\documentclass[showpacs,aps,prc,nofootinbib,showkeys,twocolumn]{revtex4-1}

\usepackage{graphicx}
\usepackage{bm}
\usepackage{amssymb}
\usepackage{amsmath,latexsym}
\usepackage[usenames]{color}
\usepackage{subfigure}
\usepackage{slashed}
\usepackage{multirow,array}
\usepackage{mathtools}
\usepackage{mathrsfs}
\usepackage[utf8]{inputenc}
\usepackage{lipsum}
\usepackage{dsfont}
\usepackage{soul}

\newcommand{\be}{\begin{equation}}
\newcommand{\ee}{\end{equation}}
\newcommand{\bs}{\begin{subequations}}
\newcommand{\es}{\end{subequations}}
\newcommand{\bea}{\begin{eqnarray}}
\newcommand{\eea}{\end{eqnarray}}

\newcommand{\G}{\text{G}}

\newcommand{\Peq}{\mathcal{P}_\text{eq}}
\newcommand{\ene}{\mathcal{E}}
\newcommand{\etas}{\eta / \mathcal{S}}
\newcommand{\tp}{\tau^\prime}
\newcommand{\tpp}{\tau^{\prime\prime}}
\newcommand{\tpi}{\tau_\pi}
\newcommand{\trel}{\tau_r}

\newcommand{\feq}{f_\text{eq}}
\newcommand{\Kn}{\text{Kn}}
\newcommand{\dt}{\partial_\tau}

\begin{document}
\title{Hydrodynamic generators in relativistic kinetic theory}
\date{\today}
\author{M.~McNelis}
\affiliation{Department of Physics, The Ohio State University, Columbus, OH 43210, USA}
\author{U.~Heinz}
\affiliation{Department of Physics, The Ohio State University, Columbus, OH 43210, USA}

\begin{abstract}
We resum the non-equilibrium gradient corrections to a single-particle distribution function evolved by the Boltzmann equation in the relaxation time approximation (RTA). We first study a system undergoing Bjorken expansion and show that, for a constant relaxation time, the exact solution of the RTA Boltzmann equation at late times (i.e. after the decay of non-hydrodynamic modes) generates the Borel resummed Chapman-Enskog series. Extending this correspondence to systems without Bjorken symmetry, we construct a (3+1)-dimensional \textit{hydrodynamic generator} for RTA kinetic theory, which is an integral representation of the Chapman-Enskog series in the limit of vanishing non-hydrodynamic modes. Relaxing this limit we find at earlier times a set of non-hydrodynamic modes coupled to the RTA Chapman-Enskog expansion. Including the dynamics of these non-hydrodynamic modes is shown to control the emergence of hydrodynamics as an effective field theory description of non-equilibrium fluids, which works well even for far-off-equilibrium situations where the Knudsen number is initially large.
\end{abstract}


\maketitle

\section{Introduction}
\label{sec1}
\vspace*{-2mm}

Viscous hydrodynamics is an effective field theory that describes the non-equilibrium physics of macroscopic systems such as liquids and gases \cite{Landau:111625,Gale:2013da,Jeon:2015dfa}. Conventional theory considers viscous hydrodynamics to be a small-gradient expansion around local equilibrium, restricting its applicability to near-equilibrium fluids \cite{Rezzolla:2013rehy}. The first-order approximation yields the Navier-Stokes equations, which is widely used in simulations of non-relativistic fluids. In the relativistic regime, the Navier-Stokes equations are acausal and therefore unstable~\cite{HISCOCK1983466}. To restore causality and stability, Israel and Stewart introduced a set of second-order relativistic relaxation equations in which the dissipative flows do not respond to the gradient forces instantaneously but relax to their Navier-Stokes solution within the slowest microscopic time scales \cite{ISRAEL1976213, ISRAEL1979341}. The decay of non-hydrodynamic modes, which are governed by microscopic processes, play an important role in the system's approach to hydrodynamics \cite{Kovtun:2005ev, Denicol:2011fa, Denicol:2012cn}. 

Causal second-order viscous hydrodynamics still assumes that the gradients of the fluid are small \cite{Denicol:2012cn}. This raises concerns about its validity when applied to relativistic fluids with large gradients \cite{Niemi:2014wta, Bazow:2016yra, Strickland:2018exs}. One could try to systematically improve the hydrodynamic equations with higher-order corrections. However, one usually finds that the hydrodynamic gradient expansion diverges, which is problematic from a theoretical standpoint \cite{Heller:2013fn, Buchel:2016cbj}. A classic example of this problem is the Chapman-Enskog expansion in kinetic theory \cite{chapman1990mathematical}. Choosing for the collision kernel the relaxation time approximation (RTA), the Boltzmann equation for the single-particle distribution function\footnote{%
    The particles are massive and on-shell unless stated otherwise.} 
$f(x,p)$ without external forces in Minkowski spacetime $x^\mu = (t,x,y,z)$ reads
\be
\label{eq:Boltzmann}
  p^\mu \partial_\mu f(x,p) = 
  \frac{p\cdot u(x)}{\tau_r(x)}
  \Bigl(\feq(x,p) - f(x,p)\Bigr)\,,
\ee
where $\feq(x,p) = \exp\left[- p\,{\cdot}\,u(x)/T(x)\right]$ is the local equilibrium distribution,\footnote{%
    Here we neglect quantum statistics and conserved charges. The degeneracy factor is set to $g = 1$.}
$u^\mu(x)$ is the fluid velocity, $T(x)$ is the temperature, and $\tau_r(x)$ is the relaxation time\footnote{%
    For simplicity, we take the relaxation time $\trel(x)$ to be momentum independent but one may replace it with $\tau_r(x,p)$.}
\cite{Anderson_Witting_1974}. The RTA Boltzmann equation can be rearranged as
\be
\label{eq:Chapman_Minkowski}
f(x,p) = \feq(x,p) - s^\mu(x,p) \partial_\mu f(x,p)\,,
\ee
where $s^\mu(x,p) = p^\mu \tau_r(x)/(p\,{\cdot}\,u(x))$. Solving this equation iteratively generates a gradient series. One obtains the first non-equilibrium gradient corrections to the distribution function after truncating the series at some finite order, establishing a connection between kinetic theory and hydrodynamics \cite{Burnett:1935, Jaiswal:2013vta}. However, a truncated approximation for the distribution function can take on unphysical negative values at sufficiently high momentum, especially if the gradients are large. As one attempts to include higher-order corrections, the series typically diverges, even for small gradients \cite{PhysRevLett.56.1571, Denicol:2016bjh}. The divergence of the Chapman-Enskog expansion is a longstanding problem in kinetic theory. Fortunately, it is known that the Chapman-Enskog expansion is asymptotic, indicating that it is Borel resummable \cite{Grad:1963}. This has been done for RTA kinetic fluids subject to Bjorken expansion, although the Borel sum has only been computed for a large but finite number of terms \cite{Heller:2016rtz, Heller:2018qvh}. Still, it is interesting to note that the Borel sum picks up a sequence of transient modes all of which have the same exponential damping factor but decay over different time periods due to their different subleading power law behavior \cite{Heller:2016rtz, Heller:2018qvh}. 

Despite these theoretical issues, causal second-order viscous hydrodynamics has proven to be a highly successful model even for far-from-equilibrium systems such as ultrarelativistic heavy-ion collisions \cite{Gale:2012rq, Shen:2014lye, Bernhard:2016tnd, Bernhard:2018hnz}. In particular, hydrodynamic simulations of small collision systems (e.g. p${+}$p collisions at the LHC), which throughout their short lifetimes feature both large longitudinal and transverse gradients, have been able to reproduce the experimentally measured anisotropic flow coefficients and other hadronic observables \cite{Shen:2016zpp, Weller:2017tsr, Heinz:2019dbd}. This unexpected string of successes gave cause for researchers to re-examine the theoretical foundations of fluid dynamics \cite{Heller:2016gbp, Romatschke:2016hle, Florkowski:2017olj, Romatschke:2017vte, Romatschke:2017ejr}. Recently, much effort has gone into understanding the so-called hydrodynamic attractor \cite{Heller:2015dha}. Various example studies on Bjorken expansion have shown that normalized hydrodynamic quantities (e.g. the temperature $\tau \partial_\tau{\ln{T}}$ and shear stress $\pi / (\ene {+} \Peq$)) with different initial conditions all evolve towards an attractor solution within a time scale on the order of the relaxation time, which is much shorter than the thermalization time \cite{Romatschke:2017vte, Romatschke:2017acs, Strickland:2017kux, Strickland:2018ayk, Jaiswal:2019cju, Chattopadhyay:2019jqj}. This strongly supports the idea that hydrodynamics can also be a valid description for far-from-equilibrium fluids \cite{Romatschke:2017vte}. However, the underlying physical mechanism that gives rise to this hydrodynamic attractor is not yet fully understood.

In this paper we resum the divergent Chapman-Enskog series of the RTA Boltzmann equation. First, we study a system subject to (0+1)-dimensional Bjorken expansion \cite{Bjorken:1982qr} where the analytic solution of the RTA Boltzmann equation is well known \cite{Baym:1984np,Florkowski:2013lya}. For a constant relaxation time, excited non-hydrodynamic modes decay completely at late times. In this case we show that the expression for the exact distribution function generates the Borel resummed Chapman-Enskog series; we interpret this integral representation of the RTA Chapman-Enskog series as a \textit{hydrodynamic generator}. We demonstrate (up to some finite order) that this correspondence also holds for Bjorken expansion with a time-dependent relaxation time, as well as for (3+1)-dimensionally expanding systems in Minkowski spacetime as long as the non-hydrodynamic modes disappear at late times. At early times, when the non-hydrodynamic modes are present, an expansion of the \textit{hydrodynamic generator} yields a Chapman-Enskog gradient expansion whose terms are initially suppressed by non-hydrodynamic modes. The decay of these non-hydrodynamic modes controls the onset of hydrodynamic behavior in non-equilibrium fluids.

\section{Chapman-Enskog expansion for Bjorken flow}
\label{chapman_enskog}
\vspace*{-2mm}

In Milne spacetime $\tilde{x}^\mu = (\tau, x, y, \eta_s)$, a transverse homogeneous system undergoing longitudinally boost-invariant Bjorken expansion is static, i.e. $\tilde{u}^\mu = (1,0,0,0)$. The RTA Boltzmann equation~\eqref{eq:Boltzmann} simplifies to
\be
\label{eq:RTA_Bjorken}
  \partial_\tau f(\tau, p) = \frac{\feq(\tau, p) - f(\tau,p)}{\trel(\tau)} \,,
\ee
where the local-equilibrium distribution is
\be
\label{eq:feq}
  \feq(\tau,p) = \exp\left[-\frac{p^\tau(\tau)}{T(\tau)}\right]\,,
\ee
with $p^\tau = \sqrt{p_\perp^2 + \tau^2(p^\eta)^2 + m^2}$. This equation can be solved analytically \cite{Baym:1984np,Florkowski:2013lya}:
\be
\label{eq:exact}
  f(\tau,p) = D(\tau,\tau_0)f_0(\tau_0,p) + {\int^\tau_{\tau_0}}{\frac{d\tp D(\tau,\tp)\feq(\tp, p)}{\trel(\tp)}} \,,
\ee
where $f_0(\tau_0,p)$ is some arbitrary initial distribution and
\be
  D(\tau_2,\tau_1) = \exp\left[-\int^{\tau_2}_{\tau_1} \frac{d\tpp}{\trel(\tpp)}\right]
\ee
is known as the damping function. One sees that the first term of the exact solution \eqref{eq:exact}, which is sensitive to the initial state, dominates the early-time dynamics. For times $\tau{\,-\,}\tau_0 \gg \trel$, however, the initial-state term decays exponentially. Hence, the second term in Eq.~\eqref{eq:exact} describes the long-time behavior of the system. 

To analyze the role that hydrodynamics plays in the evolution of this system, we turn to the Chapman-Enskog expansion. For a (0+1)-dimensional system with Bjorken symmetry, the Chapman-Enskog expansion of the RTA Boltzmann equation \eqref{eq:RTA_Bjorken} takes the form
\be
\label{eq:gradient_series}
  f_{\text{CE}}(\tau,p) = \sum_{n=0}^\infty {\left(-\trel(\tau) \partial_\tau\right)^n} \feq(\tau, p) \,.
\ee
In this gradient series, each linear operator $-\trel(\tau) \partial_\tau$ acts on all of the terms to its right. Generally, the series will contain derivatives of not only $\feq(\tau,p)$ but also $\trel(\tau)$. This causes the number of terms to grow like $n!$, which means the gradient series is divergent, even for small Knudsen numbers $\Kn \sim \trel \dt \ll 1$.\footnote{%
    Although the number of distinct gradient terms $\propto (\Kn)^n$ does not grow like $n!$, their prefactors give them the combined appearance of exhibiting $n!$ growth, assuming they have the same magnitude and sign (see for example  Eq.~\eqref{eq:fCE3}).}
One can try to resum the divergent series using Borel resummation:
\be
\label{eq:borel_sum_general}
  f^{\text B}_{\text{CE}}(\tau,p) = {\int^\infty_0} dz\, e^{-z} \sum_{n=0}^\infty \frac{{z^n}{\left({-}\trel(\tau) \partial_\tau\right)^n}\feq(\tau, p)}{n!} \,.
\ee
Here the challenge is finding a closed analytic expression for the Borel sum. Instead of computing the Borel sum directly, we analyze the exact solution \eqref{eq:exact} to look for a representation of the series. For the simplest case where the relaxation time is constant, the exact distribution function simplifies to
\be
\label{eq:exact_const_trel}
\begin{split}
  f(\tau,p) =
  & \exp\left[-\frac{(\tau - \tau_0)}{\trel}\right] f_0(\tau_0,p) \\ 
  & + \frac{1}{\trel}\int^\tau_{\tau_0} d\tp \exp\left[-\frac{(\tau - \tp)}{\trel}\right]\feq(\tp, p) \,.
\end{split}
\ee
We introduce the dimensionless coordinate $z = (\tau - \tau^\prime)\,/\,\trel$ to rewrite Eq.~\eqref{eq:exact_const_trel} as
\be
\label{eq:exact_const_trel_z}
  f(\tau,p) = e^{-z_0} f_0(\tau_0,p)\, 
  + \int^{z_0}_{0} dz \, e^{-z}\feq(\tau - \trel z, p) \,,
\ee
where $z_0 = (\tau - \tau_0)\,/\,\trel$, and Taylor expand the second term:
\be
\label{eq:exact_expand}
\begin{split}
  f(\tau,p) =
  &\  e^{-z_0} f_0(\tau_0,p) \\ 
  & + \int^{z_0}_{0} dz \, e^{-z} \sum_{n=0}^\infty \frac{(-z \trel)^n \feq^{(n)}(\tau, p)}{n!} \,,
\end{split}
\ee
with $\feq^{(n)}(\tau,p) \equiv \partial_\tau^n \feq(\tau,p)$. Sure enough, one sees that the expansion of the exact solution \eqref{eq:exact_expand} reduces to the Borel resummed Chapman-Enskog series \eqref{eq:borel_sum_general} in the limit $z_0 \to \infty$ (i.e. $\tau \to \infty$) when all non-hydrodynamic modes have decayed. With this insight, we conjecture that even for non-constant $\tau_r(\tau)$ the \textit{hydrodynamic generator}\footnote{%
    We call Eq.~\eqref{eq:general_generator} the hydrodynamic generator since it generates the hydrodynamic gradient series (7) in the limit $z_0 \to \infty$.}
\be
\label{eq:general_generator}
  f_\text{G}(\tau,p) = \int^\tau_{\tau_0}\frac{d\tp D(\tau,\tp)\feq(\tp, p)}{\trel(\tp)}
\ee
is, in the limit of vanishing non-hydrodynamic modes, an integral representation of the gradient series \eqref{eq:gradient_series}.\footnote{%
    This does not imply that the hydrodynamic generator and RTA Chapman-Enskog series are equivalent in the late time limit. The expansion of the hydrodynamic generator \eqref{eq:general_generator} may not have a finite radius of convergence.} 
If the conjecture holds it should be possible to manipulate this expression, as we did for $\tau_r{\,=\,}$const, to obtain a hydrodynamic gradient series. We use the coordinate transformation
\be
\label{eq:z_to_tau} 
  z = h(\tp, \tau) = \int^\tau_{\tp} \frac{d\tpp}{\tau_r(\tpp)}
\ee
to rewrite Eq.~\eqref{eq:general_generator} as
\be
\label{eq:generator_z}
  f_\text{G}(\tau,p) = {\int^{z_0}_0} dz \, e^{-z} \feq(h^{-1}(z,\tau), p) \,,
\ee
where 
\be
  z_0 = \int^\tau_{\tau_0} \frac{d\tpp}{\tau_r(\tpp)} \,.
\ee
Next, we compute the inverse function $\tp = h^{-1}(z,\tau)$. Physically, the relaxation time is positive and finite, which means that $z$ is a non-negative monotonic function of $\tp \in [\tau_0,\tau]$. Therefore, the function $h(\tp,\tau)$ has an inverse which we expand as a power series:\footnote{\label{fn7}%
    For a given time $\tau$, $z = h(\tp,\tau)$ is a smooth function of $\tp$ when evaluated with the exact solution~\eqref{eq:exact}; hence it can be Taylor expanded around $\tp{\,=\,}\tau$, which corresponds to a Taylor expansion of $\tau^\prime = h^{-1}(z,\tau)$ around $z{\,=\,}0.$}
%
\be
\label{eq:power_series_tau}
  \tau^\prime = h^{-1}(z,\tau) = \sum_{n=0}^\infty c_n(\tau) \, z^n \,.
\ee
The coefficients $c_n(\tau)$ can be computed by Taylor expanding Eq.~\eqref{eq:z_to_tau} around $\tp = \tau$:
\be
\label{eq:z_to_tau_series}
\begin{split}
  z &= \int^\tau_{\tp} d\tpp \sum_{n=0}^\infty \frac{(\tpp - \tau)^n}{n!} \partial_\tau^n \left[\trel^{-1}(\tau)\right] \\
  &= -\sum_{n=0}^\infty \frac{(\tp - \tau)^{n+1}}{(n+1)!} \partial_\tau^n \left[\trel^{-1}(\tau)\right] \,.
\end{split}
\ee
Inserting the power series \eqref{eq:power_series_tau} into Eq.~\eqref{eq:z_to_tau_series} we can solve for the coefficients order by order. The first coefficients are
\bs
\label{coefficients}
\begin{align}
    c_0 &= \tau \,,\\
    c_1 &= - \tau_r \,,\\
    c_2 &= \frac{\tau_r}{2!}\tau^{(1)}_r  \,,\\
    c_3 &= -\frac{\trel}{3!}\left((\tau^{(1)}_r\big)^2 + \tau_r \tau^{(2)}_r\right) \,,
\end{align}
\es
where $\tau^{(n)}_r \equiv \partial^n_\tau \tau_r(\tau)$; they satisfy the recurrence relation\footnote{%
    Using symbolic computation, we checked the validity of Eq.~\eqref{recursion} up to $n = 40$.}
\bs
\label{recursion}
\begin{align}
    c_0 &= \tau \,, \\
    c_n &= - \frac{\tau_r \partial_\tau c_{n{-}1}}{n} \indent \forall\,n\geq 1 \,.
\end{align}
\es
With these coefficients, we can now evaluate the integral \eqref{eq:generator_z} after Taylor expanding the integrand:
\be
\label{eq:fA_tau_resum}
  f_\text{G}(\tau,p) = \int^{z_0}_0 dz \, e^{-z} \sum_{n=0}^\infty \frac{(h^{-1}(z,\tau) - \tau)^n \feq^{(n)}(\tau , p)}{n!} \,.
\ee
As a demonstration, we compute the series up to $n = 3$ and truncate the expression at third order in derivatives:
\be
\label{eq:generator_truncated}
\begin{split}
  f_\text{G} \approx& \, (1 - e^{-z_0})\feq \,+\, \big(1 - \Gamma(2,z_0)\big)\delta f^{(1)} \,+\, \\
  &\Big(1 - \frac{\Gamma(3,z_0)}{2!}\Big) \delta f^{(2)} \,+\, \Big(1 - \frac{\Gamma(4,z_0)}{3!}\Big)\delta f^{(3)} \,,
\end{split}
\ee
where $\Gamma(n{+}1, z_0) = \int^\infty_{z_0} dz \, e^{-z} z^{n}$ are the upper incomplete Gamma functions. After taking the limit $z_0 \to \infty$, Eq.~\eqref{eq:generator_truncated} reduces to
\be
\label{eq:generator_truncated_limit}
  f_\text{G} \approx \feq + \delta f^{(1)} + \delta f^{(2)} + \delta f^{(3)} \,,
\ee
where 
\bs
\label{eq:fCE3}
\begin{align}
    \delta f^{(1)} =& \,- \trel \feq^{(1)} \,, \\
    \delta f^{(2)} =& \ \trel \trel^{(1)} \feq^{(1)} \,+\, \tau_r^2 \feq^{(2)} \,, \\
    \delta f^{(3)} =& -\trel \big(\trel^{(1)}\big)^2 \feq^{(1)} \,-\, \trel^2 \trel^{(2)} \feq^{(1)} \\ \nonumber
    &-\, 3 \trel^2 \trel^{(1)} \feq^{(2)} \,-\, \trel^3 \feq^{(3)} \,.
\end{align}
\es
These are precisely the non-equilibrium corrections in the Chapman-Enskog series \eqref{eq:gradient_series}. Using a computer-generated code\footnote{\label{github}%
    The codes used for this work can be downloaded at \url{https://github.com/mjmcnelis/rta_resum}.} 
we verified that the series \eqref{eq:generator_truncated} works up to order $\mathcal{O}(\Kn^{40}$):
\begin{eqnarray}
  f_\G(\tau,p) &\approx& {\int_0^{z_0}} dz \, e^{-z} \sum_{n=0}^{40} \frac{z^n{\left(-\trel(\tau) \partial_\tau\right)^n}\feq(\tau, p)}{n!} 
  \\\nonumber
  &=&\sum_{n=0}^{40} \left(1 - \frac{\Gamma(n{+}1,z_0)}{n!}\right){\left(-\trel(\tau) \partial_\tau\right)^n}\feq(\tau, p) \,.
\end{eqnarray}
This gives us a high degree of confidence that the expansion of the \textit{hydrodynamic generator}~\eqref{eq:general_generator} reduces to the Borel resummed RTA Chapman-Enskog series \eqref{eq:borel_sum_general} under the condition that the non-hydrodynamic modes decay at late times. However, at this moment we have no formal proof that this holds to all orders in the Knudsen number, due to the complexity of the expansion scheme. 

\section{Series expansion of the hydrodynamic generator}
\label{expansion}

In the limit of vanishing non-hydrodynamic modes, the \textit{hydrodynamic generator} \eqref{eq:general_generator} is an appealing representation of the RTA Chapman-Enskog series. While the Chapman-Enskog series may be divergent, the generator itself is finite, even for large Knudsen numbers. It also satisfies the RTA Boltzmann equation 
\be
\begin{split}
  \dt f_\text{G}(\tau,p) 
  =&\,\frac{\feq(\tau,p)}{\trel(\tau)} - \int^\tau_{\tau_0} \frac{d\tp D(\tau,\tp)\feq(\tp,p)}{\trel(\tau)\trel(\tp)} \\
  =&\,\frac{\feq(\tau,p) - f_\text{G}(\tau,p)}{\tau_r(\tau)}\,,
\end{split}
\ee
where we used the identities $\partial_\tau D(\tau,\tp){\,=\,}-D(\tau,\tp)/\trel(\tau)$ and $D(\tau,\tau){\,=\,}1$. However, this alone does not tell us how much hydrodynamics contributes to the dynamics of the system at finite times, before the initial state $f_0(\tau_0,p)$ has completely decayed. As long as the non-hydrodynamic modes contribute, the expansion of the exact distribution function
\be
  f(\tau,p) = e^{-z_0} f_0(\tau_0,p) + f_\text{G}(\tau,p)
\ee
around local equilibrium looks like\footnote{This expansion retains the same transseries-like structure for both early and late times. Transasymptotic solutions for the moments of the distribution function have been studied for Bjorken expansion and have been found to accurately reproduce the numerical solution of the moments equations even when continued back to earlier times~\cite{Behtash:2018moe,Behtash:2019txb}.}
\be
\label{eq:anomaly}
  f = \feq + \delta f_\G^{(0)} + \delta f_\G^{(1)} + \delta f_\G^{(2)} + \delta f_\G^{(3)} + \mathcal{O}(\Kn^4) \,,
\ee
where
\bs
\begin{align}
  \delta f_\G^{(0)} &= e^{-z_0} \left(f_0 - \feq\right) \,, \\ 
  \delta f_\G^{(n)} &= \left(1 - \frac{\Gamma(n{+}1, z_0)}{n!}\right) \delta f^{(n)} \indent \forall \, n \geq 1 \,.
\end{align}
\es
The zeroth-order correction $\delta f_\G^{(0)}$, which combines the initial-state term with the first term in Eq.~\eqref{eq:generator_truncated}, is a purely non-hydrodynamic mode and is only present for a short period of time $\sim\trel$. The other $\delta f_\G^{(n)}$ corrections are the usual hydrodynamic gradient corrections, except they are initially suppressed by their associated non-hydrodynamic mode. These non-hydrodynamic modes control the emerging strengths of the gradient corrections to the distribution function as the particle interactions drive the system towards hydrodynamics over time (i.e. as $z_0$ increases). In particular, as will be discussed below, higher-order gradient corrections are suppressed more strongly and for a longer duration than the lower-order terms.

To study these new effects on the hydrodynamic gradient expansion, we evolve a conformal fluid undergoing Bjorken expansion with the exact solution of the RTA Boltzmann equation \cite{Florkowski:2013lya, Florkowski:2013lza, Tinti:2018qfb}. We initialize the system at $\tau_0 = 0.25$ fm/$c$ with initial temperature $T(\tau_0){\,=\,}0.6$\,GeV and shear stress $\pi(\tau_0){\,=\,}0$, where $\pi{\,\equiv\,}\frac{2}{3}(\mathcal{P}_{\perp}{-}\mathcal{P}_{L})$ (by definition $\pi_\mathrm{eq}{\,=\,}0$). For the relaxation time we take $\trel{\,=\,}\tpi$ with $\tpi T = 5(\etas)$ and set the shear viscosity to entropy density ratio to $\etas = 3/(4\pi)$. Using these initial conditions we construct the temperature $T(\tau)$ by fixing the exact solution~\eqref{eq:exact} to the Landau matching condition $\ene(\tau) = 3T^4(\tau)/\pi^2$ or \cite{Florkowski:2013lya}
\be
\label{eq:T_exact}
\begin{split}
T^4(\tau) =&\, D(\tau,\tau_0) \,T^4(\tau_0)\, \mathcal{H}\left(\frac{\tau_0}{\tau}\right) \\
&+ \int_{\tau_0}^\tau \frac{d\tp}{\tau_\pi({\tp})} D(\tau,\tp) \,T^4(\tp)\, \mathcal{H}\left(\dfrac{\tp}{\tau}\right)\,,
\end{split}
\ee
where
\be
\mathcal{H}(x) = \frac{1}{2}\left(x^2 + \frac{{\tan^{-1}}\sqrt{x^{-2}-1}}{\sqrt{x^{-2}-1}}\right)\,.
\ee
The most straightforward way to solve this integral equation is by using fixed-point iteration. After computing the temperature, we evaluate the normalized shear stress\footnote{%
    This differs from the traditional definition $\bar\pi{\,\equiv\,}\pi/(\ene{+}\Peq)$ which reduces to $\bar\pi = \pi/(4\Peq)$ in the conformal limit.} 
$\bar\pi(\tau) = \pi(\tau) / \mathcal{P}_\text{eq}(\tau)$ where $\mathcal{P}_\text{eq}(\tau) = T^4(\tau)/\pi^2$ is the equilibrium pressure~\cite{Florkowski:2013lya}:
\be
\begin{split}
&\bar\pi(\tau) = D(\tau,\tau_0) \frac{T^4(\tau_0)}{T^4(\tau)}\left[\frac{1}{2}\mathcal{H}_\perp\Big(\frac{\tau_0}{\tau}\Big) -  \mathcal{H}_L\Big(\frac{\tau_0}{\tau}\Big)\right] \\
& {+} \int_{\tau_0}^\tau \frac{d\tp}{\tau_\pi({\tp})} D(\tau,\tp)\frac{T^4(\tp)}{T^4(\tau)}\left[\frac{1}{2}\mathcal{H}_\perp\Big(\frac{\tp}{\tau}\Big) -  \mathcal{H}_L\Big(\frac{\tp}{\tau}\Big)\right] \,,
\end{split}
\ee
\begin{figure}[t]
\includegraphics[width=\linewidth]{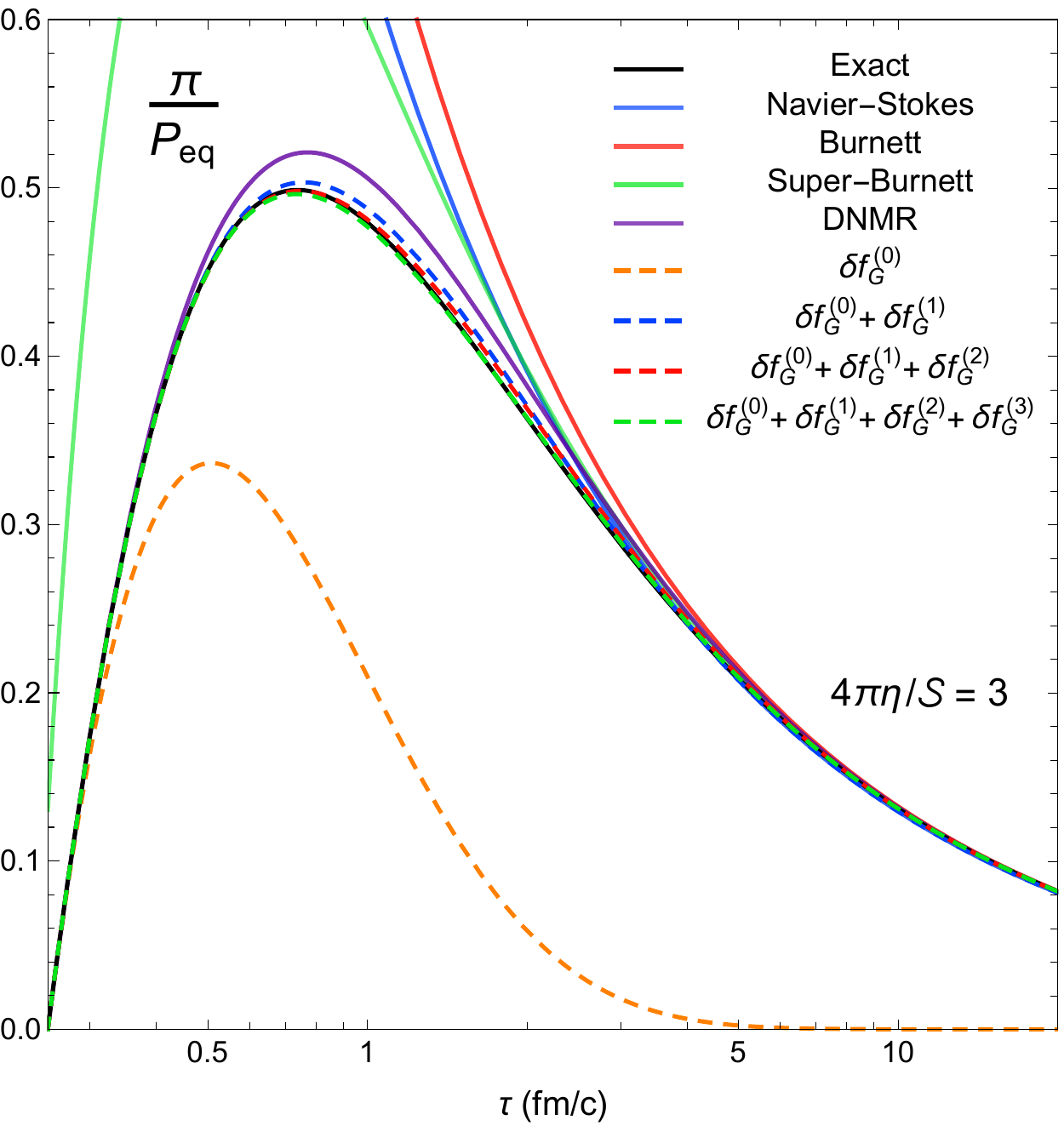}
\vspace*{-3mm}
\caption{(Color online)
\label{hydro_anomaly}
    The evolution of the pressure aniso\-tropy for conformal Bjorken expansion. The system is initialized at $\tau_0 = 0.25$ fm/$c$, with $T(\tau_0) = 0.6$\,GeV, $\pi(\tau_0) = 0$ and $\etas = 3/(4\pi)$ (see text for details). We plot the contributions of the $\delta f_\G$ corrections to the exact solution (solid black) and compare them to Navier-Stokes (solid blue), Burnett (solid red), Super-Burnett (solid green) and DNMR (solid purple) viscous hydrodynamics. As defined in footnote \ref{fn_hydrodynamization}, the system ``hydrodynamizes'' around $\tau=2.5$\,fm/$c$.
    \vspace*{-2mm}}
\end{figure}
where $\mathcal{H}_\perp$ and $\mathcal{H}_L$ are defined in App.~\ref{appb}. 

The resulting exact evolution of the normalized shear stress is shown as the solid black line in Figure~\ref{hydro_anomaly}. This exact solution is compared with various approximations discussed below. With the exact temperature and shear stress at hand, we evaluate and plot the contributions to the pressure anisotropy from the $\delta f_\G$ corrections up to third order (see Appendices \ref{appa} and \ref{appb}):
\bs
\label{eq:dfg}
\begin{align}
  \bar\pi_\G^{(0)} =&\, e^{-z_0} \frac{T^4_0}{T^4}\left[\frac{1}{2}\mathcal{H}_\perp\Big(\frac{\tau_0}{\tau}\Big) -  \mathcal{H}_L\Big(\frac{\tau_0}{\tau}\Big)\right] \,, \\
  \bar\pi_\G^{(1)} =&\, \Bigl(1 - \Gamma(2,z_0)\Bigr)\,\frac{16\tau_\pi}{15\tau} \,, \\
  \bar\pi_\G^{(2)} =&\, \Big(1 - {\textstyle\frac{1}{2!}}\Gamma(3,z_0)\Big)\,
  \frac{-16 \tau_\pi^2}{105\tau^2}\,
  \bigl(15 + 49 \tau \partial_\tau{\ln T} \bigr),\\
  \bar\pi_\G^{(3)} =&\, \Big(1 - {\textstyle\frac{1}{3!}} \Gamma(4,z_0)\Big)\, \frac{16 \tau_\pi^3}{105\tau^3} \\ \nonumber
  &\times \bigl(\tau\partial_\tau{\ln T}(135{+}182 \tau\partial_\tau{\ln T}) 
  + 77\tau^2\partial^2_\tau{\ln T} \bigr) \,.
\end{align}
\es
Here the energy conservation law and its time derivative
\bs
\label{eq:conservation_law}
\begin{align}
  \tau\partial_\tau{\ln T} &= \frac{\bar\pi - 4}{12} \,, \\
  \tau^2\partial^2_\tau{\ln T} &= \frac{4 - \bar\pi + \tau \partial_\tau \bar\pi}{12} 
\end{align}
\es
are evaluated numerically using the exact solution. We further compare these $\delta f_\G$ corrections to $\bar\pi$ to the first-order Navier-Stokes, second-order Burnett and third-order Super-Burnett solutions,
\bs
\label{eq:Navier_Burnett}
\begin{align}
\bar\pi^{(\text{NS})} &= \frac{16 \tau_\pi}{15 \tau} \,, \\
\bar\pi^{(\text{B})} &= \frac{16 \tau_\pi}{15 \tau} + \frac{64 \tau_\pi^2}{315 \tau^2} \,, \\
\bar\pi^{(\text{SB})} &= \frac{16 \tau_\pi}{15 \tau} + \frac{64 \tau_\pi^2}{315 \tau^2} - \frac{832 \tau_\pi^3}{1575 \tau^3} \,,
\end{align}
\es
as well as to the numerical solution of the causal second-order viscous hydrodynamic DNMR equations \cite{Denicol:2012cn,Denicol:2014mca}:
\bs
\label{eq:DNMR}
\begin{align}
  \tau\partial_\tau{\ln T} &= \frac{\bar\pi - 4}{12} \,, \\
  \partial_\tau \bar\pi &= -\frac{\bar\pi}{\tau_\pi} + \frac{16}{15\tau} - \frac{10\bar\pi}{21\tau} - \frac{\bar\pi^2}{3\tau} \,.
\end{align}
\es
At early times, the non-hydrodynamic mode $\delta f_\G^{(0)}$ dominates the evolution of the pressure anisotropy and is responsible for the initial rise away from the local equilibrium initial condition $\bar\pi_0=0$ (see Figure~\ref{hydro_anomaly}). As the system hydrodynamizes,\footnote{%
    \label{fn_hydrodynamization} We define hydrodynamization as the time when the leading non-hydrodynamic mode $\bar\pi_\G^{(0)}$ decays to 10\% of its maximum value. In Fig.~\ref{hydro_anomaly} this occurs at $\tau = 2.47$\,fm/$c$ (or $z_0 = 3.6$).} 
the initial-state function decays and the first-order gradient correction $\delta f^{(1)}_\G$ emerges as the leading correction to the local-equilibrium distribution $\feq$ in Eq.~\eqref{eq:anomaly}. Already, we see that the addition of $\delta f^{(1)}_\G$ nearly captures the exact pressure anisotropy. This is in stark contrast to the Navier-Stokes solution, which misses both $\delta f_\G^{(0)}$ in (\ref{eq:dfg}a) and the prefactor $1{-}\Gamma(2,z_0)$ in (\ref{eq:dfg}b) and hence fails to reproduce the shear stress for $\tau \lesssim 2$\,fm/$c$. The reader should also take note of the similarity between the blue-dashed curve and DNMR viscous hydrodynamics, where the $\delta f$ correction used to compute the transport coefficients of the relaxation equation (\ref{eq:DNMR}b) is first-order in the shear stress. 

Compared to the second-order correction accounted for in the Burnett solution (\ref{eq:Navier_Burnett}b), the full $\delta f_\G^{(2)}$ gradient correction to the shear stress is much weaker at early times since it is strongly suppressed by the corresponding non-hydrodynamic mode. By the time this non-hydrodynamic mode has decayed by 90\% (at around $\tau = 3.9$\,fm/$c$), the gradients characterized by the Knudsen number $\Kn = \tau_\pi/\tau \approx 0.2$ have already greatly diminished. As a result, the $\delta f_\G^{(2)}$ correction ends up having little overall impact on the evolution of the system. A similar observation holds for the third-order correction $\delta f_\G^{(3)}$. The combined low-order $\delta f_\G$ corrections to the local-equilibrium distribution are seen to provide excellent agreement with the exact solution; we have also checked this for different initial conditions and shear viscosities (see auxiliary materials available at the URL given in footnote \ref{github}). While we caution the reader that this does not necessarily mean the rest of the series \eqref{eq:anomaly} will converge, take this observation as justification to truncate the new expansion scheme \eqref{eq:anomaly} at a low order: Figure~\ref{hydro_anomaly} makes it clear that, at least for Bjorken flow, gradient corrections beyond first order have almost negligible influence on the fluid's dynamics during the early stages of evolution even though there the expansion rate is large. This provides a plausible explanation for the empirically observed ``unreasonable effectiveness'' \cite{Noronha-Hostler:2015wft, Heinz:2019dbd} of causal second-order viscous hydrodynamics (e.g. DNMR) even when applied outside of its conventional range of validity (e.g. when $\Kn \sim 1$).

\vspace{-2mm}
\section{Hydrodynamic generator\\ in 3+1 dimensions}
\label{hydro_generator_Minkowski}
\vspace{-2mm}

\begin{figure}[t]
\includegraphics[width=0.9\linewidth]{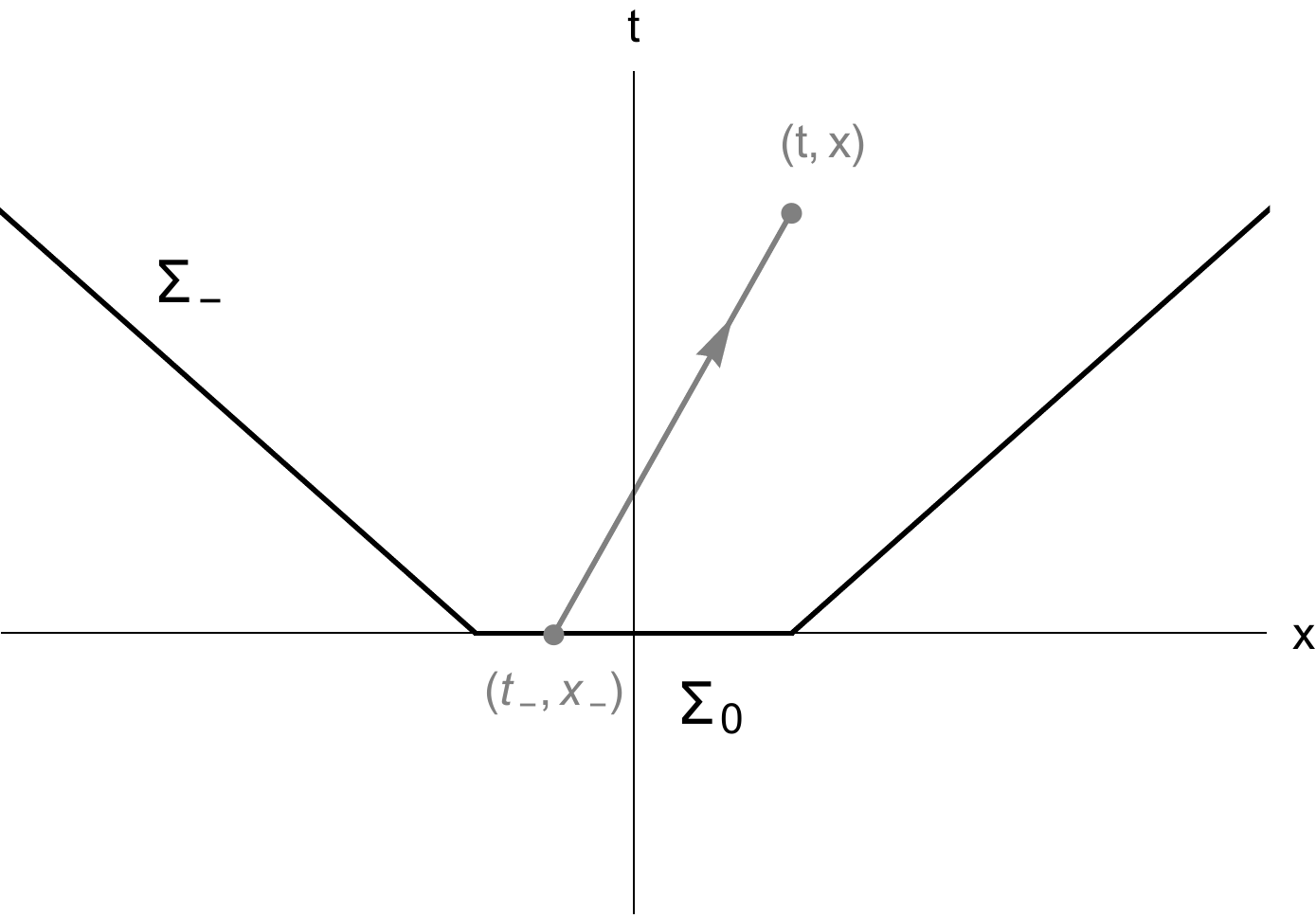}
\caption{
    \label{minkowski_diagram}
    An illustration of the path integral \eqref{eq:generator_minkowski} (solid gray) running from a point $(t_-,\bm{x}_-)$ on the hypersurface $\Sigma_-$ (solid black) to the current position $(t,\bm{x})$. The path is parallel to the particle momentum $p^\mu$ at point $(t,\bm{x})$. (Note that $x$, $x_-$ in the figure stand for 3-dimensional spatial vectors.)
}
\end{figure}

In (3+1)-dimensional Minkowski spacetime $x^\mu = (t,x,y,z)$ without Bjorken symmetry, the \textit{hydrodynamic generator} in the relaxation time approximation can be further generalized as a path integral along free-streaming characteristics:
\be
\label{eq:generator_minkowski}
  f_\text{G}(x,p) = \int^x_{x_-}{dx^\prime {\cdot\,}s^{-1}(x^\prime, p) D(x,x^\prime,p)\feq(x^\prime, p)} \,.
\ee
Here the starting point $x^\mu_- = x^\mu - (t - t_-) p^\mu / E$ lies on a hypersurface $t_- = \Sigma_-(x,y,z)$ consisting of the initial-state boundary $\Sigma_0$ and the future light cone enclosing it (see Figure \ref{minkowski_diagram}), and the reciprocal vector\footnote{%
    This formula also works for massless particles. The $p^2$ factor drops out after parametrizing the path in Eq.~\eqref{eq:generator_minkowski} as $x^{\prime\mu}(\lambda^\prime) = x^\mu - \lambda^\prime p^\mu$ ($0 \leq \lambda^\prime \leq (t - t_-) / E$) such that
    \be
    \nonumber
      dx^{\prime\mu}s^{-1}_\mu(x^\prime, p) = - \dfrac{p \cdot u(x^\prime(\lambda^\prime))}{\tau_r(x^\prime(\lambda^\prime))} \, d\lambda^\prime\,.
    \ee}
\be
\label{eq:s_inv}
  s^{-1}_\mu(x,p) = \frac{p \cdot u(x)}{\tau_r(x)} \times \frac{p_\mu}{p^2}
\ee
is constructed such that $s^\mu(x,p)\, s^{-1}_\mu(x,p) = 1$. Note that, in contrast to Eq.~(\ref{eq:general_generator}), the (3+1)-dimensional generator is not constrained by any symmetries and (after Landau matching) can accommodate any flow velocity profile $u^\mu(x)$, including ones with non-vanishing vorticity (which for Bjorken flow is forbidden by symmetry). The integral in (\ref{eq:generator_minkowski}) runs over a straight time-like characteristic line parallel to the particle momentum $p^\mu$, with a measure $dx'{\,\cdot\,}s^{-1}(x',p)$ that (unlike the one in Eq.~(\ref{eq:general_generator})) depends on momentum. The fraction of particles with momentum $p^\mu$ emitted from the thermal source $\feq(x^\prime,p)$ that travel freely through the medium and arrive at the current position $x^\mu$ unscathed is given by the damping function
%
%
\be
\label{eq:D}
  D(x,x',p) = \exp\left[-\int^x_{x'} dx^{\prime\prime}{\cdot\,}s^{-1}(x^{\prime\prime},p)\right]\,.
\ee
For short relaxation times this damping function decays very rapidly, limiting the range of influence on the fluid's dynamics at position $x$ to points $x'$ in the past light cone of $x$ with small spacetime separations $x{-}x'$.  

Let us verify that the expansion of Eq.~\eqref{eq:generator_minkowski} reduces to the more general Borel resummed RTA Chapman-Enskog expansion (up to some finite order):
\be
\label{eq:Chapman_series_Minkowski}
  f^{\text{B}}_\text{CE}(x,p) = \int_0^\infty dz\,e^{-z} \sum_{n=0}^\infty \frac{z^n(- s^\mu(x,p)\partial_\mu)^n \feq(x,p)}{n!} \,.
\ee
Following the same steps outlined in the previous section, we use the coordinate transformation
\be
\label{eq:z_minkowski}
  z = h(x^\prime, x, p) = \int^x_{x^\prime} dx^{\prime\prime}{\cdot\,}s^{-1}(x^{\prime\prime}, p)
\ee
to rewrite Eq.~\eqref{eq:generator_minkowski} as
\be
\label{eq:generator_z_minkowski}
  f_\text{G}(x,p) = \int_0^{z_-} dz\, e^{-z} \feq(h^{-1}(z, x, p), p) \,,
\ee
where
\be
  z_- = \int^x_{x_-} dx^{\prime\prime}{\cdot\,}s^{-1}(x^{\prime\prime}, p) \,.
\ee
The inverse function $h^{-1}_\mu(z, x, p)$ is now promoted to a four-vector that can be expanded as a power series:
\be
  x^\prime_\mu = h^{-1}_\mu(z, x, p) = \sum_{n=0}^\infty c_{n,\mu}(x,p)\,z^n  \,,
\ee
which can be inserted in the Taylor expansion of Eq.~\eqref{eq:z_minkowski}:
\be
\begin{split}
  z =& {\int^{\lambda^\prime}_0} d\lambda^{\prime\prime}\, g(x^{\prime\prime}(\lambda^{\prime\prime}),p) \\
  =& {\int^{\lambda^\prime}_0} d\lambda^{\prime\prime}\, \sum_{n=0}^\infty  \frac{(-\lambda^{\prime\prime})^n p^n {\cdot}\, \partial^n g(x,p)}{n!} \\
  =& -\sum_{n=0}^\infty  \frac{(-\lambda^\prime)^{n+1} p^n {\cdot}\, \partial^n g(x,p)}{(n+1)!} \,,
\end{split}
\ee
where we used the parameterization
\be
  x^{\prime\prime\mu}(\lambda^{\prime\prime}) = x^\mu - \lambda^{\prime\prime} p^\mu\,, \indent\indent (0 \leq \lambda^{\prime\prime} \leq \lambda^\prime),
\ee
with $\lambda^\prime = (x{-}x^\prime){\,\cdot\,}p \,/\, p^2$ and $g(x,p) = p{\,\cdot\,}u(x)\,/\,\tau_r(x)$. Using the product rule identities
\bs
\begin{align}
  s^\mu\partial_\nu s^{-1}_\mu =& - s^{-1}_\mu\partial_\nu s^\mu\,, \\
  s^\mu\partial_\alpha \partial_\nu s^{-1}_\mu =& - s^{-1}_\mu\partial_\alpha\partial_\nu s^\mu - (\partial_\alpha s^{-1}_\mu) (\partial_\nu s^\mu) \\ \nonumber
  &- (\partial_\nu s^{-1}_\mu) (\partial_\alpha s^\mu) \,,
\end{align}
\es
one obtains, after some algebra, the following first coefficients of the series:
\bs
\begin{align}
    c_0^{\,\mu} &= x^\mu \,, \\
    c_1^{\,\mu} &= - s^\mu \,, \\
    c_2^{\,\mu} &= \frac{1}{2!} s^\nu\partial_\nu s^\mu \,, \\
    c_3^{\,\mu} &= -\frac{1}{3!}\left((s^\alpha\partial_\alpha s^\nu)\partial_\nu s^\mu + s^\alpha s^\nu \partial_\alpha \partial_\nu s^\mu \right) \,,
\end{align}
\es
analogous to the coefficients \eqref{coefficients}. They appear to satisfy the recurrence relation
\bs
\begin{align}
     c_0^{\,\mu} &= x^\mu \,, \\
     c_n^{\,\mu} &= - \frac{s^\nu \partial_\nu c_{n{-}1}^{\,\mu}}{n} \indent \forall \, n \geq 1 \,,
\end{align}
\es
but we have not made the effort to prove this relation beyond $n{\,=\,}3$. The integral \eqref{eq:generator_z_minkowski} can then be Taylor expanded as
\be
\label{eq:generator_expand}
  f_\text{G}(x,p) {=} \int_0^{z_-}{dz\, e^{-z} \sum_{n=0}^\infty \frac{(h^{-1}(z,x,p){-}x)^n \cdot \feq^{(n)}(x, p)}{n!}} \,,
\ee
where $\feq^{(n)}(x,p) = \partial^n \feq(x,p)$. The series expansion of the \textit{hydrodynamic generator} up to third order in the Knudsen number is
\be
\begin{split}
  f_\text{G} \approx& \, (1 - e^{-z_-})\feq \,+\, \big(1 - \Gamma(2,z_-)\big)\delta f^{(1)} \,+\, \\
  &\Big(1 - \frac{\Gamma(3,z_-)}{2!}\Big) \delta f^{(2)} \,+\, \Big(1 - \frac{\Gamma(4,z_-)}{3!}\Big)\delta f^{(3)} \,,
\end{split}
\ee
which has the same structure as Eq.~(\ref{eq:generator_truncated}). For spacetime regions far in the future from the hypersurface $\Sigma_-$ in Fig.~\ref{minkowski_diagram} we assume we can take the limit $z_- \to \infty$:
\be
\label{eq:minkowski_generator_truncated}
  f_\text{G} \approx \feq + \delta f^{(1)} + \delta f^{(2)} + \delta f^{(3)} \,,
\ee
where
\bs
\begin{align}
    \delta f^{(1)} =& \,- s^\mu\partial_\mu \feq \,, \\
    \delta f^{(2)} =& \, (s^\nu\partial_\nu s^\mu) \partial_\mu \feq \,+\, s^\nu s^\mu\partial_\nu \partial_\mu \feq \,, \\
    \delta f^{(3)} =& \,- \left((s^\alpha\partial_\alpha s^\nu)\partial_\nu s^\mu\right)\partial_\mu \feq \\ \nonumber
    &\,-\, (s^\alpha s^\nu \partial_\alpha \partial_\nu s^\mu) \partial_\mu \feq \,-\, 3 s^\mu (s^\alpha \partial_\alpha s^\nu) \partial_\nu \partial_\mu \feq \\ \nonumber
    &\,-\, s^\alpha s^\nu s^\mu \partial_\alpha \partial_\nu \partial_\mu \feq \,.
\end{align}
\es
As expected, these agree with the corresponding gradient corrections from the RTA Chapman-Enskog expansion when worked out to third order. Unlike the previous section, we have not carried out the calculation \eqref{eq:generator_expand} to higher orders, because of its greater degree of complexity. We can, however, offer some reassurance by checking that the distribution function $f_\text{G}(x,p)$ given in \eqref{eq:generator_minkowski} is a particular solution of the RTA Boltzmann equation in Minkowski spacetime \eqref{eq:Chapman_Minkowski}:
\begin{eqnarray}
\label{eq:verify_fG}
  &&s^\mu(x,p) \partial_\mu f_\text{G}(x,p) 
  \nonumber\\\nonumber 
  &&= \feq(x,p) - {\int^x_{x_-}}dx^\prime {\cdot\, }s^{-1}(x^\prime, p) D(x,x^\prime,p)\feq(x^\prime, p)
  \\
  &&= \feq(x,p) - f_\text{G}(x,p) \,.
\end{eqnarray}
Here we used the identities $s^\mu(x,p)\partial_\mu D(x,x^\prime,p) {\,=\,} - D(x,x^\prime,p)$
and $D(x,x,p){\,=\,}1$.\footnote{%
    Note that the directional derivative $s^\mu \partial_\mu$ does not act on the lower limit of the path integral \eqref{eq:generator_minkowski} since the current position $x^\mu$ varies infinitesimally only along the direction of $s^\mu$, which means that the starting point $x^\mu_-$ remains fixed.}

The distribution function \eqref{eq:generator_minkowski} approaches zero on the entire initial-state surface $\Sigma_0$ since, unlike Eq.~\eqref{eq:exact}, it does not include any initial-state information. One can make use of the diagram in Fig.~\ref{minkowski_diagram} to construct and add such an initial-state term. We know that the system is initialized at time $t_0$ as $f_0(x_0,p)$, with $x^\mu_0 \in \Sigma_0$. In addition, we assume that $f_0(x_0,p)\,{=}\,0$ at the edge of $\Sigma_0$ so that $f(x,p)$ vanishes on its entire future light cone. Therefore, only characteristic lines that are connected to the initial-state surface $\Sigma_0$ as shown in Fig.~\ref{minkowski_diagram} will pick up an initial source that decays over time:
\be
  f_\text{I}(x,p) = D(x,x_-,p) f_0(x_-,p) \Theta(t_0 - t_-)\,.
\ee
Here the Heaviside step function $\Theta(t_0 {-} t_-)$ excludes those characteristic lines that end on the light cones in Fig.~\ref{minkowski_diagram} and thereby enforces $f_\text{I}(x,p) \,{=}\, 0$ if $x_-{\,\notin\,}\Sigma_0$. The full (3+1)-dimensional solution of the RTA Boltzmann equation \eqref{eq:Boltzmann} is then
\be
\label{eq:rta_solution}
  f(x,p) = f_\text{I}(x,p) + f_\text{G}(x,p)\,.
\ee
One can check that
\begin{eqnarray}
  s^\mu(x,p) \partial_\mu f(x,p) 
  &=& -f_\text{I}(x,p) + \feq(x,p) - f_\G(x,p)
\nonumber\\
  &=& \feq(x,p) - f(x,p)\,,
\end{eqnarray}
where we used the relation $s^\mu(x,p) \partial_\mu D(x,x_-,p) {\,=\,} - D(x,x_-,p)$. A more formal derivation of this solution can be found in Appendix~\ref{appc}.\footnote{%
    Eq.~\eqref{eq:rta_solution} generalizes the RTA Bjorken solution~\eqref{eq:exact} to (3+1)-dimensional systems, by replacing the integration over the fluid's history in $\tau^\prime$ with one over a path parameter $\lambda^\prime$ along a set of free-streaming past world lines. Each world line's direction depends on the momentum of the incoming particle, emitted by either an initial source $f_0(x_0,p)$ or a thermal source $\feq(x^\prime,p)$. Macroscopic observables at a given spacetime coordinate $x^\mu$ are influenced by the fluid's history encoded in these world lines.} In the free-streaming limit $\tau_r {\,\to\,} \infty$ ($z_- {\,\to\,} 0)$, the distribution function takes on the free-streaming solution $f_0(x_-,p) \Theta(t_0 - t_-)$. In the ideal hydrodynamic limit $\tau_r {\,\to\,} 0$ ($z_- {\,\to\,} \infty)$, $f(x,p) {\,\to\,} f_\G(x,p)$, which reduces to $\feq(x,p)$ since $s^\mu\partial_\mu f_\G(x,p) {\,\to\,} 0$ in Eq.~\eqref{eq:verify_fG}.\footnote{In the ideal hydrodynamic limit, the local equilibrium density operator can only accommodate flow profiles that are irrotational~\cite{Becattini:2019dxo}.}

This completes our formal argument for the RTA Boltzmann equation in the (3+1)-dimensional case. We leave its numerical implementation to future work but close this section with some thoughts about how such an implementation might look. Just like the Bjorken solution \eqref{eq:exact}, the distribution function \eqref{eq:rta_solution} for (3+1)-dimensional expansion is an {\it implicit} solution of the RTA Boltzmann equation since it depends on the temperature $T(x)$ and fluid velocity $u^\mu(x)$. In principle, the hydrodynamic fields can be reconstructed by matching the solution to the Landau frame:
\bs
\label{eq:Landau}
\begin{align}
  \mathcal{E}(x) &= \int_p (p\cdot u(x))^2 f(x,p)\,, \\
  u^\mu(x) &= \dfrac{\int_p (p\cdot u(x)) \,p^\mu f(x,p)}{\int_p (p\cdot u(x))^2 f(x,p)} \,,
\end{align}
\es
where $\int_p = \int \dfrac{d^3p}{E}$. Similar to Eq.~\eqref{eq:T_exact}, these integral equations can then be solved numerically with a root-finding algorithm such as fixed-point iteration. Starting with an approximate solution for $T(x)$ and $u^\mu(x)$, which can be provided e.g. by a viscous hydrodynamic simulation, one would repeatedly update the solution by evaluating the right-hand-side of Eq.~\eqref{eq:Landau}. Since the initial guess and exact solution share the same initial condition but may differ greatly for later times, this numerical scheme is likely to converge faster at times near $t_0$ than at later times. Instead of computing a single iteration across the entire evolution, as is commonly done \cite{Florkowski:2013lya}, it would here be more efficient to perform these iterations at a given time step until the solution is within the desired error tolerance, before proceeding to the next time step. Faster rates of convergence might be achievable if a more accurate hydrodynamic model is used to evolve the initial guess for the fluid's energy density and flow profiles. 

Due to the momentum dependence of the characteristic lines and their associated damping functions, solving the integral equations \eqref{eq:Landau} is much more involved than for the Bjorken case. Unfortunately, there does not seem to be a way of reducing the momentum-space integral without invoking additional symmetries like in Eq.~\eqref{eq:T_exact}. This leaves us with the computationally intensive task of numerically evaluating a four-dimensional integral for each spacetime point: three for the momentum and one for the path parameter along the associated characteristic line. One possible way to reduce the computing time is to parallelize at each time step the computation over the spatial grid points. Doing this on a GPU, however, still faces memory limitations: for a uniform spacetime grid, the memory required to do a full calculation of the distribution function grows rapidly with the volume of the future light cone, $V \propto t^4$. For short relaxation times, rapid damping will reduce the need for RAM to only a fraction of the fluid's evolution history. Still, the task looks formidable and will likely require a highly advanced algorithm and significant computing resources.

Based on the structure of our formal (3+1)-d solution we anticipate that effects qualitatively similar to those described in Sec.~\ref{expansion} will also be found for RTA kinetic fluids without Bjorken symmetry: at early times, the dynamics of the fluid is dominated by the non-hydrodynamic mode associated with the initial state $f_0(x_0,p)$. As time moves away from the initial-state surface $\Sigma_0$ the local-equilibrium distribution $\feq(x,p)$ and first-order gradient correction $\delta f^{(1)}(x,p)$ quickly take over, with the higher-order corrections emerging more slowly. A quantitative analysis of the contributions from the non-equilibrium corrections $\delta f_G$ to macroscopic observables will need to wait until the corresponding codes have been developed. Intermediate studies of systems with reduced symmetry (for example undergoing spherical expansion) may be useful for developing intuition and computational tools.
    
\section{Conclusions}

In this work we formulated a \textit{hydrodynamic generator} that resums the Chapman-Enskog series of the RTA Boltzmann equation. For a system with a constant relaxation time subject to Bjorken flow we have shown that the Taylor expansion of the \textit{hydrodynamic generator} reduces to the Borel resummed RTA Chapman-Enskog series in the late time limit. We then generalized the form of this \textit{hydrodynamic generator} in the relaxation time approximation to Bjorken systems with a time-dependent relaxation time, as well as to (3+1)-dimensionally expanding fluids in Minkowski spacetime without additional symmetries, outlining the methodology for reducing it to the Borel resummed RTA Chapman-Enskog series in the limit of vanishing non-hydrodynamic modes.
The mathematical proof of this correspondence to all orders in the Knudsen number is left for future work.

Our formula \eqref{eq:rta_solution} for the (3+1)-dimensional solution of the RTA Boltzmann equation in Minkowski spacetime has the nice features of being positive-definite and finite for both small and large values of the Knudsen number. It is also causal since it only depends on the present and past hydrodynamic fields. While it is not immediately obvious how to numerically implement this solution, it can potentially serve as a reference to test the validity of known viscous hydrodynamic approximations, as well as the new expansion scheme described in this work, in the relaxation time approximation without the need for Bjorken symmetry.

Most importantly, we found that the \textit{hydrodynamic generator} in RTA kinetic theory also generates a sequence of non-hydrodynamic modes that are coupled to the RTA Chapman-Enskog expansion. In RTA kinetic theory we see that these non-hydro\-dynamic modes, which decay over different time periods,\footnote{%
    We believe that these non-hydrodynamic modes are related to those identified in Refs.~\cite{Heller:2016rtz, Heller:2018qvh} using resurgence theory since the upper incomplete Gamma functions $\Gamma(n{+}1,z_0)$ have the same exponential damping $e^{-z_0}$ but different subleading polynomial factors.}
provide the mechanism that controls the emergence of hydrodynamics in non-equilibrium fluids. As the initial-state memory decays, the local-equilibrium distribution and its first-order gradient correction emerge as the leading contributors to the fluid's dynamics. Higher-order gradient corrections to the particle distribution function are suppressed during the hydrodynamization process, especially at early times. This means that even if the fluid has initially large gradients, these higher-order corrections are not as severe as traditionally thought. 

At least for systems described by the RTA Boltzmann equation discussed in this work, this extends the range of validity of causal second-order viscous hydrodynamics beyond what was traditionally assumed. It must be noted, however,\footnote{%
    We thank an anonymous referee for this comment.}
that the Boltzmann collision term (especially in the relaxation time approximation) ignores multi-particle correlations and thereby completely misses the stochastic microscopic fluctuations of the fluid and the microscopic correlations they generate \cite{Akamatsu:2016llw}. Such microscopic correlations are expected to be largest in small collision systems, such as proton-proton collisions, which also exhibit large gradients. To account for these correlations the simple damping function $D(x,x^\prime,p)$ in Eq.~(\ref{eq:D}) must be replaced by a significantly more complex Green's function, perhaps derived from a Kadanoff-Baym type equation. It remains an open question what role these stochastic microscopic fluctuations \cite{Akamatsu:2016llw, Schlichting:2019abc, Bluhm:2020mpc} play for the process of hydrodynamization in 3+1 dimensions, especially in small collision systems, and to what extent the concept of a hydrodynamic generator survives in such a more general setting.

\acknowledgements

We thank Paul Romatschke for a set of lectures delivered at Ohio State University and accompanying discussions that stimulated the work reported here. We gratefully acknowledge Chandrodoy Chattopadhyay for insightful comments on this paper and for first pointing out to us the connection between the RTA Chapman-Enskog expansion and the late-time gradient expansion of the exact Bjorken solution of the RTA Boltzmann equation via integration by parts. The numerical routine used here \cite{Tinti:2018qfb} for solving the integral equation (\ref{eq:exact}) for the exact solution of the RTA Boltzmann equation with Bjorken flow was generously provided by Gojko Vujanovic; it evolved from and improved upon an earlier code written and provided to us by Michael Strickland \cite{Florkowski:2013lya}. This work was supported by the National Science Foundation (NSF) within the framework of the JETSCAPE Collaboration under Award No. ACI-1550223. Additional partial support by the U.S. Department of Energy (DOE), Office of Science, Office for Nuclear Physics under Award No. DE-SC0004286 and within the framework of the BEST and JET Collaborations is also acknowledged.

\begin{appendix}
\section{Standard gradient corrections}
\label{appa}
Here we compute the standard gradient corrections to the normalized shear stress for a conformal system undergoing Bjorken expansion:
\be
  \bar\pi = \bar\pi^{(1)} + \bar\pi^{(2)} + \bar\pi^{(3)} + \mathcal{O}(\Kn^4) \,.
\ee
Before proceeding, we make a change in variables $w = \tau^2 p^\eta$ to rewrite the local-equilibrium distribution~\eqref{eq:feq} as
\be
\feq(\tau,p_T,w) = \exp\left[-\frac{\sqrt{\tau^2p_\perp^2 + w^2}}{\tau T(\tau)}\right] \,.
\ee
The first-order gradient correction to the distribution function is
\be
\label{eq:df1}
\begin{split}
\delta f^{(1)} &= - \trel \partial_\tau \feq \\
&= - \frac{\tpi \feq}{\tau^2 T} \left(\frac{w^2}{v} + v \tau \partial_\tau \ln T\right) \,,
\end{split}
\ee
where we set $\trel \,{=}\, \tau_\pi \,{=}\, 5(\etas)/T$ (taking $\etas$ as a constant) and $v = \sqrt{\tau^2p_\perp^2 + w^2}$. The first-order correction to the shear stress is then
\be
\label{eq:dPL1}
\bar\pi^{(1)} = \frac{2\pi^2}{3T^4} \int_p \left(\frac{p_\perp^2}{2} - \frac{w^2}{\tau^2}\right) \delta f^{(1)} \,,
\ee
where $\int_p = \int \dfrac{d^2 p_\perp dw}{v(2\pi)^3}$. After inserting $\delta f^{(1)}$ in Eq.~\eqref{eq:dPL1} and substituting the spherical coordinates
\bs
\begin{align}
\tau p^x &= v \sin\theta \cos\phi \,, \\
\tau p^y &= v \sin\theta \sin\phi \,, \\
w &= v \cos\theta \,,
\end{align}
\es
one obtains
\be
\bar\pi^{(1)} = \frac{16\tpi}{15\tau} \,.
\ee
The $\delta f^{(2)}$ and $\delta f^{(3)}$ corrections are too cumbersome to list here. We simply state the results for the second and third-order shear corrections (for the derivation see the auxiliary materials available in the github repository referenced in footnote~\ref{github}):
\bs
\label{eq:pi23}
\begin{align}
\bar\pi^{(2)} &= - \frac{16 \tau_\pi^2}{105\tau^2} \left(15 + 49 \tau \partial_\tau{\ln T} \right) \,, \\
\bar\pi^{(3)} &= \frac{16 \tau_\pi^3}{105\tau^3} \left(\tau\partial_\tau{\ln T}(135 {+} 182\tau\partial_\tau{\ln T}) + 77\tau^2\partial^2_\tau{\ln T} \right) \,.
\end{align}
\es
Here we used the relation $\partial_\tau \tau_\pi = - \tau_\pi \partial_\tau{\ln T}$ to eliminate time derivatives of the shear relaxation time. At late times, the gradients $\Kn = \tau_\pi / \tau  \sim \tau^{-2/3}$ become small. Hence, the asymptotic solutions for the energy conservation law and its time derivative \eqref{eq:conservation_law} are
\bs
\begin{align}
  \tau \, \partial_\tau{\ln T} &= -\frac{1}{3} + \frac{4\tpi}{45\tau} + \mathcal{O}\left(\Kn^2\right) \,, \\
  \tau^2 \partial^2_\tau{\ln T} &= \frac{1}{3} + \mathcal{O}\left(\Kn\right) \,.
\end{align}
\es
The shear corrections \eqref{eq:pi23} then reduce to
\bs
\label{eq:shear_asymptotic}
\begin{align}
  \bar\pi^{(2)} &\approx \frac{64 \tau_\pi^2}{315\tau^2} - \frac{448 \tau_\pi^3}{675\tau^3} \,,\\
  \bar\pi^{(3)} &\approx \frac{128 \tau_\pi^3}{945\tau^3} \,.
\end{align}
\es
Finally, the non-hydrodynamic modes in the exact solution \eqref{eq:anomaly} decay at late times since $z_0 \sim \tau^{2/3}$. Using this, the second and third-order corrections in Eq.~\eqref{eq:shear_asymptotic} can be regrouped as
\be
  \bar\pi^{(2)} \to \frac{64 \tau_\pi^2}{315\tau^2} \,, \qquad
  \bar\pi^{(3)} \to -\frac{832 \tau_\pi^3}{1575\tau^3} \,,
\ee
which appear in the Burnett and Super-Burnett solutions (\ref{eq:Navier_Burnett}b-c).

\section{Leading non-hydrodynamic mode correction}
\label{appb}

Here we compute the shear stress correction from the leading non-hydrodynamic mode $\delta f_\G^{(0)} = e^{-z_0}(f_0 - \feq)$:
\be
\label{eq:shear_nonhydro}
  \bar\pi^{(0)}_\G = \frac{2\pi^2 e^{-z_0}}{3T^4} \int_p \left(\frac{p_\perp^2}{2} - \frac{w^2}{\tau^2}\right) f_0(\tau_0,p_T,w) \,,
\ee
where the second term $\propto \feq(\tau,p_T,w)$ vanishes by symmetry. The code \cite{Florkowski:2013lya, Florkowski:2013lza,Tinti:2018qfb} that evolves the RTA Bjorken solution \eqref{eq:exact} gives the user the option to initialize the distribution as~\cite{Florkowski:2013lya,Florkowski:2013lza}
\be
  f_0(\tau_0, p_T, w) = \exp\left[-\frac{\sqrt{\tau_0^2p_\perp^2 + (1+\xi_0)w^2}}{\tau_0 \Lambda_0}\right] \,,
\ee
where
\be
\Lambda_0 = T_0 \, \mathcal{H}\big((1+\xi_0)^{-1/2}\big)^{-1/4}
\ee
is the effective temperature and $\xi_0$ is an anisotropy parameter that deforms the longitudinal momentum space. After substituting the spherical coordinates
\bs
\begin{align}
\tau_0 p^x &= v_0 \sin\theta \cos\phi \,, \\
\tau_0 p^y &= v_0 \sin\theta \sin\phi \,, \\
(1+\xi_0)^{1/2}w &= v_0 \cos\theta \,,
\end{align}
\es
where $v_0 = \sqrt{\tau_0^2p_\perp^2 + (1+\xi_0)w^2}$, Eq.~\eqref{eq:shear_nonhydro} can be rewritten as
\be
\label{eq:pi0_aniso}
\bar\pi^{(0)}_\G = e^{-z_0}\frac{\Lambda^4_0}{T^4}\left[\frac{1}{2}\mathcal{H}_\perp\Big(\frac{\tau_0 \alpha_{L0}}{\tau}\Big) {-}  \mathcal{H}_L\Big(\frac{\tau_0\alpha_{L0}}{\tau}\Big)\right] \,,
\ee
where $\alpha_{L0} = (1+\xi_0)^{-1/2}$ and the hypergeometric functions \cite{Florkowski:2013lya} 
\bs
\begin{align}
\mathcal{H}_\perp(x) &=\, x \int_{{-}1}^1 \frac{d{\cos\theta} \, (1 - \cos^2\theta)}{\sqrt{1 + (x^2{-}1)\cos^2\theta}} \,, \\
\mathcal{H}_L(x) &=\, x^3 \int_{{-}1}^1 \frac{d{\cos\theta} \, \cos^2\theta}{\sqrt{1 + (x^2{-}1)\cos^2\theta}}
\end{align}
\es
are 
\bs
\begin{align}
  \mathcal{H}_\perp(x) &=\, \frac{1}{1{-}x^2}\big(x^2 + (1{-}2x^2)\, \mathcal{T}(x^{-2}{-}1)\big) \,, \\ 
  \mathcal{H}_L(x) &=\,\frac{x^2}{1{-}x^2}\big({-}x^2 + \mathcal{T}(x^{-2}{-}1)\big) \,,
\end{align}
\es
with $\mathcal{T}(y) = \dfrac{\tan^{-1}{\sqrt{y}}}{\sqrt{y}}$. In Sec.~\ref{expansion}, we had initialized the shear stress to $\pi(\tau_0) = 0$ so that $\xi_0 = 0$ and $\Lambda_0 = T_0$. Then Eq.~\eqref{eq:pi0_aniso} reduces to Eq. (\ref{eq:dfg}a).

\section{Solution of the RTA Boltzmann equation in 3+1 dimensions}
\label{appc}

In this Appendix we derive the (3+1)-dimensional solution of the RTA Boltzmann equation in Minkowski spacetime. First, we rewrite Eq.~\eqref{eq:Boltzmann} as
\be
\label{eq:RTA_C}
  s^\mu(x,p)\partial_\mu f(x,p) + f(x,p) = \feq(x,p)
\ee
and multiply both sides by the function
\be
\label{eq:C2}
  q(x,p) = \exp\left[{\int_{x_\star}^x} dx^{\prime\prime} {\cdot\, } s^{-1}(x^{\prime\prime},p)\right] \,,
\ee
where the path integral runs over a straight line that is parallel to $p^\mu$; the coordinate $x_\star^\mu$ is a fixed point on the characteristic line (see Fig.~\ref{minkowski_diagram})
\be
\label{eq:characteristic}
{x^\prime}^\mu(\lambda^\prime) = x^\mu - \lambda^\prime p^\mu  \indent \indent 0 \leq \lambda^\prime \leq \lambda_- \,,
\ee
with $\lambda_- = (t - t_-) / E$. Eq.~\eqref{eq:RTA_C} can be rewritten as
\be
s^\mu(x,p)\partial_\mu \left[q(x,p) f(x,p)\right] = q(x,p) \feq(x,p)\,. 
\ee
Now we integrate this equation along the characteristic line~\eqref{eq:characteristic}:
\be
\label{eq:integrate_character}
\begin{split}
&\int_{x_-}^x dx^\prime {\cdot\, } s^{-1}(x^\prime,p)\, s^\mu(x^\prime,p)\frac{\partial\left[q(x^\prime,p) f(x^\prime,p)\right]}{\partial {x^\prime}^\mu} \\
&= \int_{x_-}^x dx^\prime {\cdot\, } s^{-1}(x^\prime,p)\, q(x^\prime,p) \feq(x^\prime,p) \,.
\end{split}
\ee
The left-hand-side of Eq.~\eqref{eq:integrate_character} can be parameterized in terms of $\lambda^\prime$:
\be
\begin{split}
&\int_{\lambda_-}^0 d\lambda^\prime \frac{d\left[q(x^\prime(\lambda^\prime),p) f(x^\prime(\lambda^\prime),p)\right]}{d\lambda^\prime} \\
&= q(x,p) f(x,p) - q(x_-,p) f(x_-,p) \,,
\end{split}
\ee
where we used the relations $d{x^\prime}^\nu = -p^\nu d\lambda^\prime$ and $\dfrac{\partial}{\partial {x^\prime}^\mu} = -\dfrac{p_\mu}{p^2} \dfrac{d}{d\lambda^\prime}$. For the distribution function on the hypersurface $\Sigma_-$ (see Fig.~\ref{minkowski_diagram}) we take
\be
f(x_-,p) = f_0(x_-,p) \Theta(t_0 - t_-) \,.
\ee
The solution of the RTA Boltzmann equation is then
\be
\begin{split}
\label{eq:C8}
  f(x,p) =& \,\frac{q(x_-,p)}{q(x,p)} f_0(x_-,p) \Theta(t_0 - t_-) \\
  &+ \int_{x_-}^x dx^\prime {\cdot\, } s^{-1}(x^\prime,p)\, \frac{q(x^\prime,p)}{q(x,p)} \feq(x^\prime,p) \,.
\end{split}
\ee
After using the identity $D(x_2,x_1,p) = q(x_1,p) \,/\, q(x_2,p)$, one arrives at Eq.~\eqref{eq:rta_solution}.
\end{appendix}

\bibliography{generator}

\begin{thebibliography}{55}%
\makeatletter
\providecommand \@ifxundefined [1]{%
 \@ifx{#1\undefined}
}%
\providecommand \@ifnum [1]{%
 \ifnum #1\expandafter \@firstoftwo
 \else \expandafter \@secondoftwo
 \fi
}%
\providecommand \@ifx [1]{%
 \ifx #1\expandafter \@firstoftwo
 \else \expandafter \@secondoftwo
 \fi
}%
\providecommand \natexlab [1]{#1}%
\providecommand \enquote  [1]{``#1''}%
\providecommand \bibnamefont  [1]{#1}%
\providecommand \bibfnamefont [1]{#1}%
\providecommand \citenamefont [1]{#1}%
\providecommand \href@noop [0]{\@secondoftwo}%
\providecommand \href [0]{\begingroup \@sanitize@url \@href}%
\providecommand \@href[1]{\@@startlink{#1}\@@href}%
\providecommand \@@href[1]{\endgroup#1\@@endlink}%
\providecommand \@sanitize@url [0]{\catcode `\\12\catcode `\$12\catcode
  `\&12\catcode `\#12\catcode `\^12\catcode `\_12\catcode `\%12\relax}%
\providecommand \@@startlink[1]{}%
\providecommand \@@endlink[0]{}%
\providecommand \url  [0]{\begingroup\@sanitize@url \@url }%
\providecommand \@url [1]{\endgroup\@href {#1}{\urlprefix }}%
\providecommand \urlprefix  [0]{URL }%
\providecommand \Eprint [0]{\href }%
\providecommand \doibase [0]{http://dx.doi.org/}%
\providecommand \selectlanguage [0]{\@gobble}%
\providecommand \bibinfo  [0]{\@secondoftwo}%
\providecommand \bibfield  [0]{\@secondoftwo}%
\providecommand \translation [1]{[#1]}%
\providecommand \BibitemOpen [0]{}%
\providecommand \bibitemStop [0]{}%
\providecommand \bibitemNoStop [0]{.\EOS\space}%
\providecommand \EOS [0]{\spacefactor3000\relax}%
\providecommand \BibitemShut  [1]{\csname bibitem#1\endcsname}%
\let\auto@bib@innerbib\@empty
\bibitem [{\citenamefont {Landau}\ and\ \citenamefont
  {Lifshitz}(1987)}]{Landau:111625}%
  \BibitemOpen
  \bibfield  {author} {\bibinfo {author} {\bibfnamefont {L.~D.}\ \bibnamefont
  {Landau}}\ and\ \bibinfo {author} {\bibfnamefont {E.~M.}\ \bibnamefont
  {Lifshitz}},\ }\href {https://cds.cern.ch/record/111625} {\emph {\bibinfo
  {title} {{Fluid mechanics; 2nd ed.}}}},\ Course of theoretical physics\
  (\bibinfo  {publisher} {Butterworth},\ \bibinfo {address} {Oxford},\ \bibinfo
  {year} {1987})\BibitemShut {NoStop}%
\bibitem [{\citenamefont {Gale}\ \emph
  {et~al.}(2013{\natexlab{a}})\citenamefont {Gale}, \citenamefont {Jeon},\ and\
  \citenamefont {Schenke}}]{Gale:2013da}%
  \BibitemOpen
  \bibfield  {author} {\bibinfo {author} {\bibfnamefont {C.}~\bibnamefont
  {Gale}}, \bibinfo {author} {\bibfnamefont {S.}~\bibnamefont {Jeon}}, \ and\
  \bibinfo {author} {\bibfnamefont {B.}~\bibnamefont {Schenke}},\ }\href
  {\doibase 10.1142/S0217751X13400113} {\bibfield  {journal} {\bibinfo
  {journal} {Int. J. Mod. Phys.}\ }\textbf {\bibinfo {volume} {A28}},\ \bibinfo
  {pages} {1340011} (\bibinfo {year} {2013}{\natexlab{a}})},\ \Eprint
  {http://arxiv.org/abs/1301.5893} {arXiv:1301.5893 [nucl-th]} \BibitemShut
  {NoStop}%
\bibitem [{\citenamefont {Jeon}\ and\ \citenamefont
  {Heinz}(2015)}]{Jeon:2015dfa}%
  \BibitemOpen
  \bibfield  {author} {\bibinfo {author} {\bibfnamefont {S.}~\bibnamefont
  {Jeon}}\ and\ \bibinfo {author} {\bibfnamefont {U.}~\bibnamefont {Heinz}},\
  }\href {\doibase 10.1142/S0218301315300106} {\bibfield  {journal} {\bibinfo
  {journal} {Int. J. Mod. Phys.}\ }\textbf {\bibinfo {volume} {E24}},\ \bibinfo
  {pages} {1530010} (\bibinfo {year} {2015})},\ \Eprint
  {http://arxiv.org/abs/1503.03931} {arXiv:1503.03931 [hep-ph]} \BibitemShut
  {NoStop}%
\bibitem [{\citenamefont {{Rezzolla}}\ and\ \citenamefont
  {{Zanotti}}(2013)}]{Rezzolla:2013rehy}%
  \BibitemOpen
  \bibfield  {author} {\bibinfo {author} {\bibfnamefont {L.}~\bibnamefont
  {{Rezzolla}}}\ and\ \bibinfo {author} {\bibfnamefont {O.}~\bibnamefont
  {{Zanotti}}},\ }\href {\doibase 10.1093/acprof:oso/9780198528906.001.0001}
  {\emph {\bibinfo {title} {{Relativistic Hydrodynamics}}}}\ (\bibinfo
  {publisher} {Oxford University Press, London},\ \bibinfo {year}
  {2013})\BibitemShut {NoStop}%
\bibitem [{\citenamefont {Hiscock}\ and\ \citenamefont
  {Lindblom}(1983)}]{HISCOCK1983466}%
  \BibitemOpen
  \bibfield  {author} {\bibinfo {author} {\bibfnamefont {W.~A.}\ \bibnamefont
  {Hiscock}}\ and\ \bibinfo {author} {\bibfnamefont {L.}~\bibnamefont
  {Lindblom}},\ }\href {\doibase https://doi.org/10.1016/0003-4916(83)90288-9}
  {\bibfield  {journal} {\bibinfo  {journal} {Annals of Physics}\ }\textbf
  {\bibinfo {volume} {151}},\ \bibinfo {pages} {466 } (\bibinfo {year}
  {1983})}\BibitemShut {NoStop}%
\bibitem [{\citenamefont {Israel}\ and\ \citenamefont
  {Stewart}(1976)}]{ISRAEL1976213}%
  \BibitemOpen
  \bibfield  {author} {\bibinfo {author} {\bibfnamefont {W.}~\bibnamefont
  {Israel}}\ and\ \bibinfo {author} {\bibfnamefont {J.}~\bibnamefont
  {Stewart}},\ }\href {\doibase https://doi.org/10.1016/0375-9601(76)90075-X}
  {\bibfield  {journal} {\bibinfo  {journal} {Physics Letters A}\ }\textbf
  {\bibinfo {volume} {58}},\ \bibinfo {pages} {213 } (\bibinfo {year}
  {1976})}\BibitemShut {NoStop}%
\bibitem [{\citenamefont {Israel}\ and\ \citenamefont
  {Stewart}(1979)}]{ISRAEL1979341}%
  \BibitemOpen
  \bibfield  {author} {\bibinfo {author} {\bibfnamefont {W.}~\bibnamefont
  {Israel}}\ and\ \bibinfo {author} {\bibfnamefont {J.}~\bibnamefont
  {Stewart}},\ }\href {\doibase https://doi.org/10.1016/0003-4916(79)90130-1}
  {\bibfield  {journal} {\bibinfo  {journal} {Annals of Physics}\ }\textbf
  {\bibinfo {volume} {118}},\ \bibinfo {pages} {341 } (\bibinfo {year}
  {1979})}\BibitemShut {NoStop}%
\bibitem [{\citenamefont {Kovtun}\ and\ \citenamefont
  {Starinets}(2005)}]{Kovtun:2005ev}%
  \BibitemOpen
  \bibfield  {author} {\bibinfo {author} {\bibfnamefont {P.~K.}\ \bibnamefont
  {Kovtun}}\ and\ \bibinfo {author} {\bibfnamefont {A.~O.}\ \bibnamefont
  {Starinets}},\ }\href {\doibase 10.1103/PhysRevD.72.086009} {\bibfield
  {journal} {\bibinfo  {journal} {Phys. Rev.}\ }\textbf {\bibinfo {volume}
  {D72}},\ \bibinfo {pages} {086009} (\bibinfo {year} {2005})},\ \Eprint
  {http://arxiv.org/abs/hep-th/0506184} {arXiv:hep-th/0506184 [hep-th]}
  \BibitemShut {NoStop}%
\bibitem [{\citenamefont {Denicol}\ \emph {et~al.}(2011)\citenamefont
  {Denicol}, \citenamefont {Noronha}, \citenamefont {Niemi},\ and\
  \citenamefont {Rischke}}]{Denicol:2011fa}%
  \BibitemOpen
  \bibfield  {author} {\bibinfo {author} {\bibfnamefont {G.~S.}\ \bibnamefont
  {Denicol}}, \bibinfo {author} {\bibfnamefont {J.}~\bibnamefont {Noronha}},
  \bibinfo {author} {\bibfnamefont {H.}~\bibnamefont {Niemi}}, \ and\ \bibinfo
  {author} {\bibfnamefont {D.~H.}\ \bibnamefont {Rischke}},\ }\href {\doibase
  10.1103/PhysRevD.83.074019} {\bibfield  {journal} {\bibinfo  {journal} {Phys.
  Rev.}\ }\textbf {\bibinfo {volume} {D83}},\ \bibinfo {pages} {074019}
  (\bibinfo {year} {2011})},\ \Eprint {http://arxiv.org/abs/1102.4780}
  {arXiv:1102.4780 [hep-th]} \BibitemShut {NoStop}%
\bibitem [{\citenamefont {Denicol}\ \emph {et~al.}(2012)\citenamefont
  {Denicol}, \citenamefont {Niemi}, \citenamefont {Molnar},\ and\ \citenamefont
  {Rischke}}]{Denicol:2012cn}%
  \BibitemOpen
  \bibfield  {author} {\bibinfo {author} {\bibfnamefont {G.~S.}\ \bibnamefont
  {Denicol}}, \bibinfo {author} {\bibfnamefont {H.}~\bibnamefont {Niemi}},
  \bibinfo {author} {\bibfnamefont {E.}~\bibnamefont {Molnar}}, \ and\ \bibinfo
  {author} {\bibfnamefont {D.~H.}\ \bibnamefont {Rischke}},\ }\href {\doibase
  10.1103/PhysRevD.85.114047, 10.1103/PhysRevD.91.039902} {\bibfield  {journal}
  {\bibinfo  {journal} {Phys. Rev.}\ }\textbf {\bibinfo {volume} {D85}},\
  \bibinfo {pages} {114047} (\bibinfo {year} {2012})},\ \bibinfo {note}
  {[Erratum: Phys. Rev.D91,no.3,039902(2015)]},\ \Eprint
  {http://arxiv.org/abs/1202.4551} {arXiv:1202.4551 [nucl-th]} \BibitemShut
  {NoStop}%
\bibitem [{\citenamefont {Niemi}\ and\ \citenamefont
  {Denicol}(2014)}]{Niemi:2014wta}%
  \BibitemOpen
  \bibfield  {author} {\bibinfo {author} {\bibfnamefont {H.}~\bibnamefont
  {Niemi}}\ and\ \bibinfo {author} {\bibfnamefont {G.~S.}\ \bibnamefont
  {Denicol}},\ }\href@noop {} {\  (\bibinfo {year} {2014})},\ \Eprint
  {http://arxiv.org/abs/1404.7327} {arXiv:1404.7327 [nucl-th]} \BibitemShut
  {NoStop}%
\bibitem [{\citenamefont {Bazow}\ \emph {et~al.}(2018)\citenamefont {Bazow},
  \citenamefont {Heinz},\ and\ \citenamefont {Strickland}}]{Bazow:2016yra}%
  \BibitemOpen
  \bibfield  {author} {\bibinfo {author} {\bibfnamefont {D.}~\bibnamefont
  {Bazow}}, \bibinfo {author} {\bibfnamefont {U.}~\bibnamefont {Heinz}}, \ and\
  \bibinfo {author} {\bibfnamefont {M.}~\bibnamefont {Strickland}},\ }\href
  {\doibase 10.1016/j.cpc.2017.01.015} {\bibfield  {journal} {\bibinfo
  {journal} {Comput. Phys. Commun.}\ }\textbf {\bibinfo {volume} {225}},\
  \bibinfo {pages} {92} (\bibinfo {year} {2018})},\ \Eprint
  {http://arxiv.org/abs/1608.06577} {arXiv:1608.06577 [physics.comp-ph]}
  \BibitemShut {NoStop}%
\bibitem [{\citenamefont {Strickland}(2019)}]{Strickland:2018exs}%
  \BibitemOpen
  \bibfield  {author} {\bibinfo {author} {\bibfnamefont {M.}~\bibnamefont
  {Strickland}},\ }\href {\doibase 10.1016/j.nuclphysa.2018.09.071} {\bibfield
  {journal} {\bibinfo  {journal} {Nucl. Phys.}\ }\textbf {\bibinfo {volume}
  {A982}},\ \bibinfo {pages} {92} (\bibinfo {year} {2019})},\ \Eprint
  {http://arxiv.org/abs/1807.07191} {arXiv:1807.07191 [nucl-th]} \BibitemShut
  {NoStop}%
\bibitem [{\citenamefont {Heller}\ \emph {et~al.}(2013)\citenamefont {Heller},
  \citenamefont {Janik},\ and\ \citenamefont {Witaszczyk}}]{Heller:2013fn}%
  \BibitemOpen
  \bibfield  {author} {\bibinfo {author} {\bibfnamefont {M.~P.}\ \bibnamefont
  {Heller}}, \bibinfo {author} {\bibfnamefont {R.~A.}\ \bibnamefont {Janik}}, \
  and\ \bibinfo {author} {\bibfnamefont {P.}~\bibnamefont {Witaszczyk}},\
  }\href {\doibase 10.1103/PhysRevLett.110.211602} {\bibfield  {journal}
  {\bibinfo  {journal} {Phys. Rev. Lett.}\ }\textbf {\bibinfo {volume} {110}},\
  \bibinfo {pages} {211602} (\bibinfo {year} {2013})},\ \Eprint
  {http://arxiv.org/abs/1302.0697} {arXiv:1302.0697 [hep-th]} \BibitemShut
  {NoStop}%
\bibitem [{\citenamefont {Buchel}\ \emph {et~al.}(2016)\citenamefont {Buchel},
  \citenamefont {Heller},\ and\ \citenamefont {Noronha}}]{Buchel:2016cbj}%
  \BibitemOpen
  \bibfield  {author} {\bibinfo {author} {\bibfnamefont {A.}~\bibnamefont
  {Buchel}}, \bibinfo {author} {\bibfnamefont {M.~P.}\ \bibnamefont {Heller}},
  \ and\ \bibinfo {author} {\bibfnamefont {J.}~\bibnamefont {Noronha}},\ }\href
  {\doibase 10.1103/PhysRevD.94.106011} {\bibfield  {journal} {\bibinfo
  {journal} {Phys. Rev.}\ }\textbf {\bibinfo {volume} {D94}},\ \bibinfo {pages}
  {106011} (\bibinfo {year} {2016})},\ \Eprint
  {http://arxiv.org/abs/1603.05344} {arXiv:1603.05344 [hep-th]} \BibitemShut
  {NoStop}%
\bibitem [{\citenamefont {Chapman}\ \emph {et~al.}(1990)\citenamefont
  {Chapman}, \citenamefont {Cowling}, \citenamefont {Burnett},\ and\
  \citenamefont {Cercignani}}]{chapman1990mathematical}%
  \BibitemOpen
  \bibfield  {author} {\bibinfo {author} {\bibfnamefont {S.}~\bibnamefont
  {Chapman}}, \bibinfo {author} {\bibfnamefont {T.}~\bibnamefont {Cowling}},
  \bibinfo {author} {\bibfnamefont {D.}~\bibnamefont {Burnett}}, \ and\
  \bibinfo {author} {\bibfnamefont {C.}~\bibnamefont {Cercignani}},\ }\href
  {https://books.google.com/books?id=Cbp5JP2OTrwC} {\emph {\bibinfo {title}
  {The Mathematical Theory of Non-uniform Gases: An Account of the Kinetic
  Theory of Viscosity, Thermal Conduction and Diffusion in Gases}}},\ Cambridge
  Mathematical Library\ (\bibinfo  {publisher} {Cambridge University Press},\
  \bibinfo {year} {1990})\BibitemShut {NoStop}%
\bibitem [{\citenamefont {Anderson}\ and\ \citenamefont
  {Witting}(1974)}]{Anderson_Witting_1974}%
  \BibitemOpen
  \bibfield  {author} {\bibinfo {author} {\bibfnamefont {J.}~\bibnamefont
  {Anderson}}\ and\ \bibinfo {author} {\bibfnamefont {H.}~\bibnamefont
  {Witting}},\ }\href@noop {} {\bibfield  {journal} {\bibinfo  {journal}
  {Physica}\ }\textbf {\bibinfo {volume} {74}},\ \bibinfo {pages} {466}
  (\bibinfo {year} {1974})}\BibitemShut {NoStop}%
\bibitem [{\citenamefont {Burnett}(1935)}]{Burnett:1935}%
  \BibitemOpen
  \bibfield  {author} {\bibinfo {author} {\bibfnamefont {D.}~\bibnamefont
  {Burnett}},\ }\href {\doibase 10.1112/plms/s2-39.1.385} {\bibfield  {journal}
  {\bibinfo  {journal} {Proceedings of the London Mathematical Society}\
  }\textbf {\bibinfo {volume} {s2-39}},\ \bibinfo {pages} {385} (\bibinfo
  {year} {1935})}\BibitemShut {NoStop}%
\bibitem [{\citenamefont {Jaiswal}(2013)}]{Jaiswal:2013vta}%
  \BibitemOpen
  \bibfield  {author} {\bibinfo {author} {\bibfnamefont {A.}~\bibnamefont
  {Jaiswal}},\ }\href {\doibase 10.1103/PhysRevC.88.021903} {\bibfield
  {journal} {\bibinfo  {journal} {Phys. Rev.}\ }\textbf {\bibinfo {volume}
  {C88}},\ \bibinfo {pages} {021903(R)} (\bibinfo {year} {2013})},\ \Eprint
  {http://arxiv.org/abs/1305.3480} {arXiv:1305.3480 [nucl-th]} \BibitemShut
  {NoStop}%
\bibitem [{\citenamefont {Santos}\ \emph {et~al.}(1986)\citenamefont {Santos},
  \citenamefont {Brey},\ and\ \citenamefont {Dufty}}]{PhysRevLett.56.1571}%
  \BibitemOpen
  \bibfield  {author} {\bibinfo {author} {\bibfnamefont {A.}~\bibnamefont
  {Santos}}, \bibinfo {author} {\bibfnamefont {J.~J.}\ \bibnamefont {Brey}}, \
  and\ \bibinfo {author} {\bibfnamefont {J.~W.}\ \bibnamefont {Dufty}},\ }\href
  {\doibase 10.1103/PhysRevLett.56.1571} {\bibfield  {journal} {\bibinfo
  {journal} {Phys. Rev. Lett.}\ }\textbf {\bibinfo {volume} {56}},\ \bibinfo
  {pages} {1571} (\bibinfo {year} {1986})}\BibitemShut {NoStop}%
\bibitem [{\citenamefont {Denicol}\ and\ \citenamefont
  {Noronha}(2016)}]{Denicol:2016bjh}%
  \BibitemOpen
  \bibfield  {author} {\bibinfo {author} {\bibfnamefont {G.~S.}\ \bibnamefont
  {Denicol}}\ and\ \bibinfo {author} {\bibfnamefont {J.}~\bibnamefont
  {Noronha}},\ }\href@noop {} {\  (\bibinfo {year} {2016})},\ \Eprint
  {http://arxiv.org/abs/1608.07869} {arXiv:1608.07869 [nucl-th]} \BibitemShut
  {NoStop}%
\bibitem [{\citenamefont {{Grad}}(1963)}]{Grad:1963}%
  \BibitemOpen
  \bibfield  {author} {\bibinfo {author} {\bibfnamefont {H.}~\bibnamefont
  {{Grad}}},\ }\href {\doibase 10.1063/1.1706716} {\bibfield  {journal}
  {\bibinfo  {journal} {Physics of Fluids}\ }\textbf {\bibinfo {volume} {6}},\
  \bibinfo {pages} {147} (\bibinfo {year} {1963})}\BibitemShut {NoStop}%
\bibitem [{\citenamefont {Heller}\ \emph {et~al.}(2018)\citenamefont {Heller},
  \citenamefont {Kurkela}, \citenamefont {Spaliński},\ and\ \citenamefont
  {Svensson}}]{Heller:2016rtz}%
  \BibitemOpen
  \bibfield  {author} {\bibinfo {author} {\bibfnamefont {M.~P.}\ \bibnamefont
  {Heller}}, \bibinfo {author} {\bibfnamefont {A.}~\bibnamefont {Kurkela}},
  \bibinfo {author} {\bibfnamefont {M.}~\bibnamefont {Spaliński}}, \ and\
  \bibinfo {author} {\bibfnamefont {V.}~\bibnamefont {Svensson}},\ }\href
  {\doibase 10.1103/PhysRevD.97.091503} {\bibfield  {journal} {\bibinfo
  {journal} {Phys. Rev.}\ }\textbf {\bibinfo {volume} {D97}},\ \bibinfo {pages}
  {091503(R)} (\bibinfo {year} {2018})},\ \Eprint
  {http://arxiv.org/abs/1609.04803} {arXiv:1609.04803 [nucl-th]} \BibitemShut
  {NoStop}%
\bibitem [{\citenamefont {Heller}\ and\ \citenamefont
  {Svensson}(2018)}]{Heller:2018qvh}%
  \BibitemOpen
  \bibfield  {author} {\bibinfo {author} {\bibfnamefont {M.~P.}\ \bibnamefont
  {Heller}}\ and\ \bibinfo {author} {\bibfnamefont {V.}~\bibnamefont
  {Svensson}},\ }\href {\doibase 10.1103/PhysRevD.98.054016} {\bibfield
  {journal} {\bibinfo  {journal} {Phys. Rev.}\ }\textbf {\bibinfo {volume}
  {D98}},\ \bibinfo {pages} {054016} (\bibinfo {year} {2018})},\ \Eprint
  {http://arxiv.org/abs/1802.08225} {arXiv:1802.08225 [nucl-th]} \BibitemShut
  {NoStop}%
\bibitem [{\citenamefont {Gale}\ \emph
  {et~al.}(2013{\natexlab{b}})\citenamefont {Gale}, \citenamefont {Jeon},
  \citenamefont {Schenke}, \citenamefont {Tribedy},\ and\ \citenamefont
  {Venugopalan}}]{Gale:2012rq}%
  \BibitemOpen
  \bibfield  {author} {\bibinfo {author} {\bibfnamefont {C.}~\bibnamefont
  {Gale}}, \bibinfo {author} {\bibfnamefont {S.}~\bibnamefont {Jeon}}, \bibinfo
  {author} {\bibfnamefont {B.}~\bibnamefont {Schenke}}, \bibinfo {author}
  {\bibfnamefont {P.}~\bibnamefont {Tribedy}}, \ and\ \bibinfo {author}
  {\bibfnamefont {R.}~\bibnamefont {Venugopalan}},\ }\href {\doibase
  10.1103/PhysRevLett.110.012302} {\bibfield  {journal} {\bibinfo  {journal}
  {Phys. Rev. Lett.}\ }\textbf {\bibinfo {volume} {110}},\ \bibinfo {pages}
  {012302} (\bibinfo {year} {2013}{\natexlab{b}})},\ \Eprint
  {http://arxiv.org/abs/1209.6330} {arXiv:1209.6330 [nucl-th]} \BibitemShut
  {NoStop}%
\bibitem [{\citenamefont {Shen}(2014)}]{Shen:2014lye}%
  \BibitemOpen
  \bibfield  {author} {\bibinfo {author} {\bibfnamefont {C.}~\bibnamefont
  {Shen}},\ }\emph {\bibinfo {title} {{The standard model for relativistic
  heavy-ion collisions and electromagnetic tomography}}},\ \href@noop {} {Ph.D.
  thesis},\ \bibinfo  {school} {Ohio State U.} (\bibinfo {year}
  {2014})\BibitemShut {NoStop}%
\bibitem [{\citenamefont {Bernhard}\ \emph {et~al.}(2016)\citenamefont
  {Bernhard}, \citenamefont {Moreland}, \citenamefont {Bass}, \citenamefont
  {Liu},\ and\ \citenamefont {Heinz}}]{Bernhard:2016tnd}%
  \BibitemOpen
  \bibfield  {author} {\bibinfo {author} {\bibfnamefont {J.~E.}\ \bibnamefont
  {Bernhard}}, \bibinfo {author} {\bibfnamefont {J.~S.}\ \bibnamefont
  {Moreland}}, \bibinfo {author} {\bibfnamefont {S.~A.}\ \bibnamefont {Bass}},
  \bibinfo {author} {\bibfnamefont {J.}~\bibnamefont {Liu}}, \ and\ \bibinfo
  {author} {\bibfnamefont {U.}~\bibnamefont {Heinz}},\ }\href {\doibase
  10.1103/PhysRevC.94.024907} {\bibfield  {journal} {\bibinfo  {journal} {Phys.
  Rev.}\ }\textbf {\bibinfo {volume} {C94}},\ \bibinfo {pages} {024907}
  (\bibinfo {year} {2016})},\ \Eprint {http://arxiv.org/abs/1605.03954}
  {arXiv:1605.03954 [nucl-th]} \BibitemShut {NoStop}%
\bibitem [{\citenamefont {Bernhard}(2018)}]{Bernhard:2018hnz}%
  \BibitemOpen
  \bibfield  {author} {\bibinfo {author} {\bibfnamefont {J.~E.}\ \bibnamefont
  {Bernhard}},\ }\emph {\bibinfo {title} {{Bayesian parameter estimation for
  relativistic heavy-ion collisions}}},\ \href@noop {} {Ph.D. thesis},\
  \bibinfo  {school} {Duke U.} (\bibinfo {year} {2018}),\ \Eprint
  {http://arxiv.org/abs/1804.06469} {arXiv:1804.06469 [nucl-th]} \BibitemShut
  {NoStop}%
\bibitem [{\citenamefont {Shen}\ \emph {et~al.}(2017)\citenamefont {Shen},
  \citenamefont {Paquet}, \citenamefont {Denicol}, \citenamefont {Jeon},\ and\
  \citenamefont {Gale}}]{Shen:2016zpp}%
  \BibitemOpen
  \bibfield  {author} {\bibinfo {author} {\bibfnamefont {C.}~\bibnamefont
  {Shen}}, \bibinfo {author} {\bibfnamefont {J.-F.}\ \bibnamefont {Paquet}},
  \bibinfo {author} {\bibfnamefont {G.~S.}\ \bibnamefont {Denicol}}, \bibinfo
  {author} {\bibfnamefont {S.}~\bibnamefont {Jeon}}, \ and\ \bibinfo {author}
  {\bibfnamefont {C.}~\bibnamefont {Gale}},\ }\href {\doibase
  10.1103/PhysRevC.95.014906} {\bibfield  {journal} {\bibinfo  {journal} {Phys.
  Rev.}\ }\textbf {\bibinfo {volume} {C95}},\ \bibinfo {pages} {014906}
  (\bibinfo {year} {2017})},\ \Eprint {http://arxiv.org/abs/1609.02590}
  {arXiv:1609.02590 [nucl-th]} \BibitemShut {NoStop}%
\bibitem [{\citenamefont {Weller}\ and\ \citenamefont
  {Romatschke}(2017)}]{Weller:2017tsr}%
  \BibitemOpen
  \bibfield  {author} {\bibinfo {author} {\bibfnamefont {R.~D.}\ \bibnamefont
  {Weller}}\ and\ \bibinfo {author} {\bibfnamefont {P.}~\bibnamefont
  {Romatschke}},\ }\href {\doibase 10.1016/j.physletb.2017.09.077} {\bibfield
  {journal} {\bibinfo  {journal} {Phys. Lett.}\ }\textbf {\bibinfo {volume}
  {B774}},\ \bibinfo {pages} {351} (\bibinfo {year} {2017})},\ \Eprint
  {http://arxiv.org/abs/1701.07145} {arXiv:1701.07145 [nucl-th]} \BibitemShut
  {NoStop}%
\bibitem [{\citenamefont {Heinz}\ and\ \citenamefont
  {Moreland}(2019)}]{Heinz:2019dbd}%
  \BibitemOpen
  \bibfield  {author} {\bibinfo {author} {\bibfnamefont {U.}~\bibnamefont
  {Heinz}}\ and\ \bibinfo {author} {\bibfnamefont {J.~S.}\ \bibnamefont
  {Moreland}},\ }\href {\doibase 10.1088/1742-6596/1271/1/012018} {\bibfield
  {journal} {\bibinfo  {journal} {J. Phys. Conf. Ser.}\ }\textbf {\bibinfo
  {volume} {1271}},\ \bibinfo {pages} {012018} (\bibinfo {year} {2019})},\
  \Eprint {http://arxiv.org/abs/1904.06592} {arXiv:1904.06592 [nucl-th]}
  \BibitemShut {NoStop}%
\bibitem [{\citenamefont {Heller}(2016)}]{Heller:2016gbp}%
  \BibitemOpen
  \bibfield  {author} {\bibinfo {author} {\bibfnamefont {M.~P.}\ \bibnamefont
  {Heller}},\ }\href {\doibase 10.5506/APhysPolB.47.2581} {\bibfield  {journal}
  {\bibinfo  {journal} {Acta Phys. Polon.}\ }\textbf {\bibinfo {volume}
  {B47}},\ \bibinfo {pages} {2581} (\bibinfo {year} {2016})},\ \Eprint
  {http://arxiv.org/abs/1610.02023} {arXiv:1610.02023 [hep-th]} \BibitemShut
  {NoStop}%
\bibitem [{\citenamefont
  {Romatschke}(2017{\natexlab{a}})}]{Romatschke:2016hle}%
  \BibitemOpen
  \bibfield  {author} {\bibinfo {author} {\bibfnamefont {P.}~\bibnamefont
  {Romatschke}},\ }\href {\doibase 10.1140/epjc/s10052-016-4567-x} {\bibfield
  {journal} {\bibinfo  {journal} {Eur. Phys. J.}\ }\textbf {\bibinfo {volume}
  {C77}},\ \bibinfo {pages} {21} (\bibinfo {year} {2017}{\natexlab{a}})},\
  \Eprint {http://arxiv.org/abs/1609.02820} {arXiv:1609.02820 [nucl-th]}
  \BibitemShut {NoStop}%
\bibitem [{\citenamefont {Florkowski}\ \emph {et~al.}(2018)\citenamefont
  {Florkowski}, \citenamefont {Heller},\ and\ \citenamefont
  {Spalinski}}]{Florkowski:2017olj}%
  \BibitemOpen
  \bibfield  {author} {\bibinfo {author} {\bibfnamefont {W.}~\bibnamefont
  {Florkowski}}, \bibinfo {author} {\bibfnamefont {M.~P.}\ \bibnamefont
  {Heller}}, \ and\ \bibinfo {author} {\bibfnamefont {M.}~\bibnamefont
  {Spalinski}},\ }\href {\doibase 10.1088/1361-6633/aaa091} {\bibfield
  {journal} {\bibinfo  {journal} {Rept. Prog. Phys.}\ }\textbf {\bibinfo
  {volume} {81}},\ \bibinfo {pages} {046001} (\bibinfo {year} {2018})},\
  \Eprint {http://arxiv.org/abs/1707.02282} {arXiv:1707.02282 [hep-ph]}
  \BibitemShut {NoStop}%
\bibitem [{\citenamefont {Romatschke}(2018)}]{Romatschke:2017vte}%
  \BibitemOpen
  \bibfield  {author} {\bibinfo {author} {\bibfnamefont {P.}~\bibnamefont
  {Romatschke}},\ }\href {\doibase 10.1103/PhysRevLett.120.012301} {\bibfield
  {journal} {\bibinfo  {journal} {Phys. Rev. Lett.}\ }\textbf {\bibinfo
  {volume} {120}},\ \bibinfo {pages} {012301} (\bibinfo {year} {2018})},\
  \Eprint {http://arxiv.org/abs/1704.08699} {arXiv:1704.08699 [hep-th]}
  \BibitemShut {NoStop}%
\bibitem [{\citenamefont {Romatschke}\ and\ \citenamefont
  {Romatschke}(2019)}]{Romatschke:2017ejr}%
  \BibitemOpen
  \bibfield  {author} {\bibinfo {author} {\bibfnamefont {P.}~\bibnamefont
  {Romatschke}}\ and\ \bibinfo {author} {\bibfnamefont {U.}~\bibnamefont
  {Romatschke}},\ }\href {\doibase 10.1017/9781108651998} {\emph {\bibinfo
  {title} {{Relativistic fluid dynamics in and out of equilibrium}}}},\
  Cambridge Monographs on Mathematical Physics\ (\bibinfo  {publisher}
  {Cambridge University Press},\ \bibinfo {year} {2019})\ \Eprint
  {http://arxiv.org/abs/1712.05815} {arXiv:1712.05815 [nucl-th]} \BibitemShut
  {NoStop}%
\bibitem [{\citenamefont {Heller}\ and\ \citenamefont
  {Spalinski}(2015)}]{Heller:2015dha}%
  \BibitemOpen
  \bibfield  {author} {\bibinfo {author} {\bibfnamefont {M.~P.}\ \bibnamefont
  {Heller}}\ and\ \bibinfo {author} {\bibfnamefont {M.}~\bibnamefont
  {Spalinski}},\ }\href {\doibase 10.1103/PhysRevLett.115.072501} {\bibfield
  {journal} {\bibinfo  {journal} {Phys. Rev. Lett.}\ }\textbf {\bibinfo
  {volume} {115}},\ \bibinfo {pages} {072501} (\bibinfo {year} {2015})},\
  \Eprint {http://arxiv.org/abs/1503.07514} {arXiv:1503.07514 [hep-th]}
  \BibitemShut {NoStop}%
\bibitem [{\citenamefont
  {Romatschke}(2017{\natexlab{b}})}]{Romatschke:2017acs}%
  \BibitemOpen
  \bibfield  {author} {\bibinfo {author} {\bibfnamefont {P.}~\bibnamefont
  {Romatschke}},\ }\href {\doibase 10.1007/JHEP12(2017)079} {\bibfield
  {journal} {\bibinfo  {journal} {JHEP}\ }\textbf {\bibinfo {volume} {12}},\
  \bibinfo {pages} {079} (\bibinfo {year} {2017}{\natexlab{b}})},\ \Eprint
  {http://arxiv.org/abs/1710.03234} {arXiv:1710.03234 [hep-th]} \BibitemShut
  {NoStop}%
\bibitem [{\citenamefont {Strickland}\ \emph {et~al.}(2018)\citenamefont
  {Strickland}, \citenamefont {Noronha},\ and\ \citenamefont
  {Denicol}}]{Strickland:2017kux}%
  \BibitemOpen
  \bibfield  {author} {\bibinfo {author} {\bibfnamefont {M.}~\bibnamefont
  {Strickland}}, \bibinfo {author} {\bibfnamefont {J.}~\bibnamefont {Noronha}},
  \ and\ \bibinfo {author} {\bibfnamefont {G.~S.}\ \bibnamefont {Denicol}},\
  }\href {\doibase 10.1103/PhysRevD.97.036020} {\bibfield  {journal} {\bibinfo
  {journal} {Phys. Rev.}\ }\textbf {\bibinfo {volume} {D97}},\ \bibinfo {pages}
  {036020} (\bibinfo {year} {2018})},\ \Eprint
  {http://arxiv.org/abs/1709.06644} {arXiv:1709.06644 [nucl-th]} \BibitemShut
  {NoStop}%
\bibitem [{\citenamefont {Strickland}(2018)}]{Strickland:2018ayk}%
  \BibitemOpen
  \bibfield  {author} {\bibinfo {author} {\bibfnamefont {M.}~\bibnamefont
  {Strickland}},\ }\href {\doibase 10.1007/JHEP12(2018)128} {\bibfield
  {journal} {\bibinfo  {journal} {JHEP}\ }\textbf {\bibinfo {volume} {12}},\
  \bibinfo {pages} {128} (\bibinfo {year} {2018})},\ \Eprint
  {http://arxiv.org/abs/1809.01200} {arXiv:1809.01200 [nucl-th]} \BibitemShut
  {NoStop}%
\bibitem [{\citenamefont {Jaiswal}\ \emph {et~al.}(2019)\citenamefont
  {Jaiswal}, \citenamefont {Chattopadhyay}, \citenamefont {Jaiswal},
  \citenamefont {Pal},\ and\ \citenamefont {Heinz}}]{Jaiswal:2019cju}%
  \BibitemOpen
  \bibfield  {author} {\bibinfo {author} {\bibfnamefont {S.}~\bibnamefont
  {Jaiswal}}, \bibinfo {author} {\bibfnamefont {C.}~\bibnamefont
  {Chattopadhyay}}, \bibinfo {author} {\bibfnamefont {A.}~\bibnamefont
  {Jaiswal}}, \bibinfo {author} {\bibfnamefont {S.}~\bibnamefont {Pal}}, \ and\
  \bibinfo {author} {\bibfnamefont {U.}~\bibnamefont {Heinz}},\ }\href
  {\doibase 10.1103/PhysRevC.100.034901} {\bibfield  {journal} {\bibinfo
  {journal} {Phys. Rev.}\ }\textbf {\bibinfo {volume} {C100}},\ \bibinfo
  {pages} {034901} (\bibinfo {year} {2019})},\ \Eprint
  {http://arxiv.org/abs/1907.07965} {arXiv:1907.07965 [nucl-th]} \BibitemShut
  {NoStop}%
\bibitem [{\citenamefont {Chattopadhyay}\ and\ \citenamefont
  {Heinz}(2020)}]{Chattopadhyay:2019jqj}%
  \BibitemOpen
  \bibfield  {author} {\bibinfo {author} {\bibfnamefont {C.}~\bibnamefont
  {Chattopadhyay}}\ and\ \bibinfo {author} {\bibfnamefont {U.}~\bibnamefont
  {Heinz}},\ }\href {\doibase 10.1016/j.physletb.2019.135158} {\bibfield
  {journal} {\bibinfo  {journal} {Phys. Lett.}\ }\textbf {\bibinfo {volume}
  {B801}},\ \bibinfo {pages} {135158} (\bibinfo {year} {2020})},\ \Eprint
  {http://arxiv.org/abs/1911.07765} {arXiv:1911.07765 [nucl-th]} \BibitemShut
  {NoStop}%
\bibitem [{\citenamefont {Bjorken}(1983)}]{Bjorken:1982qr}%
  \BibitemOpen
  \bibfield  {author} {\bibinfo {author} {\bibfnamefont {J.~D.}\ \bibnamefont
  {Bjorken}},\ }\href {\doibase 10.1103/PhysRevD.27.140} {\bibfield  {journal}
  {\bibinfo  {journal} {Phys. Rev.}\ }\textbf {\bibinfo {volume} {D27}},\
  \bibinfo {pages} {140} (\bibinfo {year} {1983})}\BibitemShut {NoStop}%
\bibitem [{\citenamefont {Baym}(1984)}]{Baym:1984np}%
  \BibitemOpen
  \bibfield  {author} {\bibinfo {author} {\bibfnamefont {G.}~\bibnamefont
  {Baym}},\ }\href {\doibase 10.1016/0370-2693(84)91863-X} {\bibfield
  {journal} {\bibinfo  {journal} {Phys. Lett.}\ }\textbf {\bibinfo {volume}
  {138B}},\ \bibinfo {pages} {18} (\bibinfo {year} {1984})}\BibitemShut
  {NoStop}%
\bibitem [{\citenamefont {Florkowski}\ \emph
  {et~al.}(2013{\natexlab{a}})\citenamefont {Florkowski}, \citenamefont
  {Ryblewski},\ and\ \citenamefont {Strickland}}]{Florkowski:2013lya}%
  \BibitemOpen
  \bibfield  {author} {\bibinfo {author} {\bibfnamefont {W.}~\bibnamefont
  {Florkowski}}, \bibinfo {author} {\bibfnamefont {R.}~\bibnamefont
  {Ryblewski}}, \ and\ \bibinfo {author} {\bibfnamefont {M.}~\bibnamefont
  {Strickland}},\ }\href {\doibase 10.1103/PhysRevC.88.024903} {\bibfield
  {journal} {\bibinfo  {journal} {Phys. Rev.}\ }\textbf {\bibinfo {volume}
  {C88}},\ \bibinfo {pages} {024903} (\bibinfo {year} {2013}{\natexlab{a}})},\
  \Eprint {http://arxiv.org/abs/1305.7234} {arXiv:1305.7234 [nucl-th]}
  \BibitemShut {NoStop}%
\bibitem [{\citenamefont {Behtash}\ \emph
  {et~al.}(2019{\natexlab{a}})\citenamefont {Behtash}, \citenamefont
  {Cruz-Camacho}, \citenamefont {Kamata},\ and\ \citenamefont
  {Martinez}}]{Behtash:2018moe}%
  \BibitemOpen
  \bibfield  {author} {\bibinfo {author} {\bibfnamefont {A.}~\bibnamefont
  {Behtash}}, \bibinfo {author} {\bibfnamefont {C.~N.}\ \bibnamefont
  {Cruz-Camacho}}, \bibinfo {author} {\bibfnamefont {S.}~\bibnamefont
  {Kamata}}, \ and\ \bibinfo {author} {\bibfnamefont {M.}~\bibnamefont
  {Martinez}},\ }\href {\doibase 10.1016/j.physletb.2019.134914} {\bibfield
  {journal} {\bibinfo  {journal} {Phys. Lett.}\ }\textbf {\bibinfo {volume}
  {B797}},\ \bibinfo {pages} {134914} (\bibinfo {year} {2019}{\natexlab{a}})},\
  \Eprint {http://arxiv.org/abs/1805.07881} {arXiv:1805.07881 [hep-th]}
  \BibitemShut {NoStop}%
\bibitem [{\citenamefont {Behtash}\ \emph
  {et~al.}(2019{\natexlab{b}})\citenamefont {Behtash}, \citenamefont {Kamata},
  \citenamefont {Martinez},\ and\ \citenamefont {Shi}}]{Behtash:2019txb}%
  \BibitemOpen
  \bibfield  {author} {\bibinfo {author} {\bibfnamefont {A.}~\bibnamefont
  {Behtash}}, \bibinfo {author} {\bibfnamefont {S.}~\bibnamefont {Kamata}},
  \bibinfo {author} {\bibfnamefont {M.}~\bibnamefont {Martinez}}, \ and\
  \bibinfo {author} {\bibfnamefont {H.}~\bibnamefont {Shi}},\ }\href {\doibase
  10.1103/PhysRevD.99.116012} {\bibfield  {journal} {\bibinfo  {journal} {Phys.
  Rev.}\ }\textbf {\bibinfo {volume} {D99}},\ \bibinfo {pages} {116012}
  (\bibinfo {year} {2019}{\natexlab{b}})},\ \Eprint
  {http://arxiv.org/abs/1901.08632} {arXiv:1901.08632 [hep-th]} \BibitemShut
  {NoStop}%
\bibitem [{\citenamefont {Florkowski}\ \emph
  {et~al.}(2013{\natexlab{b}})\citenamefont {Florkowski}, \citenamefont
  {Ryblewski},\ and\ \citenamefont {Strickland}}]{Florkowski:2013lza}%
  \BibitemOpen
  \bibfield  {author} {\bibinfo {author} {\bibfnamefont {W.}~\bibnamefont
  {Florkowski}}, \bibinfo {author} {\bibfnamefont {R.}~\bibnamefont
  {Ryblewski}}, \ and\ \bibinfo {author} {\bibfnamefont {M.}~\bibnamefont
  {Strickland}},\ }\href {\doibase 10.1016/j.nuclphysa.2013.08.004} {\bibfield
  {journal} {\bibinfo  {journal} {Nucl. Phys.}\ }\textbf {\bibinfo {volume}
  {A916}},\ \bibinfo {pages} {249} (\bibinfo {year} {2013}{\natexlab{b}})},\
  \Eprint {http://arxiv.org/abs/1304.0665} {arXiv:1304.0665 [nucl-th]}
  \BibitemShut {NoStop}%
\bibitem [{\citenamefont {Tinti}\ \emph {et~al.}(2019)\citenamefont {Tinti},
  \citenamefont {Vujanovic}, \citenamefont {Noronha},\ and\ \citenamefont
  {Heinz}}]{Tinti:2018qfb}%
  \BibitemOpen
  \bibfield  {author} {\bibinfo {author} {\bibfnamefont {L.}~\bibnamefont
  {Tinti}}, \bibinfo {author} {\bibfnamefont {G.}~\bibnamefont {Vujanovic}},
  \bibinfo {author} {\bibfnamefont {J.}~\bibnamefont {Noronha}}, \ and\
  \bibinfo {author} {\bibfnamefont {U.}~\bibnamefont {Heinz}},\ }\href
  {\doibase 10.1103/PhysRevD.99.016009} {\bibfield  {journal} {\bibinfo
  {journal} {Phys. Rev.}\ }\textbf {\bibinfo {volume} {D99}},\ \bibinfo {pages}
  {016009} (\bibinfo {year} {2019})},\ \bibinfo {note}
  {\url{https://github.com/gvujan/Boltzmann_equation_solver_w_RTA_and_Bjorken_sym}},\
  \Eprint {http://arxiv.org/abs/1808.06436} {arXiv:1808.06436 [nucl-th]}
  \BibitemShut {NoStop}%
\bibitem [{\citenamefont {Denicol}\ \emph {et~al.}(2014)\citenamefont
  {Denicol}, \citenamefont {Florkowski}, \citenamefont {Ryblewski},\ and\
  \citenamefont {Strickland}}]{Denicol:2014mca}%
  \BibitemOpen
  \bibfield  {author} {\bibinfo {author} {\bibfnamefont {G.~S.}\ \bibnamefont
  {Denicol}}, \bibinfo {author} {\bibfnamefont {W.}~\bibnamefont {Florkowski}},
  \bibinfo {author} {\bibfnamefont {R.}~\bibnamefont {Ryblewski}}, \ and\
  \bibinfo {author} {\bibfnamefont {M.}~\bibnamefont {Strickland}},\ }\href
  {\doibase 10.1103/PhysRevC.90.044905} {\bibfield  {journal} {\bibinfo
  {journal} {Phys. Rev.}\ }\textbf {\bibinfo {volume} {C90}},\ \bibinfo {pages}
  {044905} (\bibinfo {year} {2014})},\ \Eprint {http://arxiv.org/abs/1407.4767}
  {arXiv:1407.4767 [hep-ph]} \BibitemShut {NoStop}%
\bibitem [{\citenamefont {Noronha-Hostler}\ \emph {et~al.}(2016)\citenamefont
  {Noronha-Hostler}, \citenamefont {Noronha},\ and\ \citenamefont
  {Gyulassy}}]{Noronha-Hostler:2015wft}%
  \BibitemOpen
  \bibfield  {author} {\bibinfo {author} {\bibfnamefont {J.}~\bibnamefont
  {Noronha-Hostler}}, \bibinfo {author} {\bibfnamefont {J.}~\bibnamefont
  {Noronha}}, \ and\ \bibinfo {author} {\bibfnamefont {M.}~\bibnamefont
  {Gyulassy}},\ }\href {\doibase 10.1016/j.nuclphysa.2016.01.050} {\bibfield
  {journal} {\bibinfo  {journal} {Nucl. Phys. A}\ }\textbf {\bibinfo {volume}
  {956}},\ \bibinfo {pages} {890} (\bibinfo {year} {2016})},\ \Eprint
  {http://arxiv.org/abs/1512.07135} {arXiv:1512.07135 [nucl-th]} \BibitemShut
  {NoStop}%
\bibitem [{\citenamefont {Becattini}\ \emph {et~al.}(2019)\citenamefont
  {Becattini}, \citenamefont {Buzzegoli},\ and\ \citenamefont
  {Grossi}}]{Becattini:2019dxo}%
  \BibitemOpen
  \bibfield  {author} {\bibinfo {author} {\bibfnamefont {F.}~\bibnamefont
  {Becattini}}, \bibinfo {author} {\bibfnamefont {M.}~\bibnamefont
  {Buzzegoli}}, \ and\ \bibinfo {author} {\bibfnamefont {E.}~\bibnamefont
  {Grossi}},\ }\href {\doibase 10.3390/particles2020014} {\bibfield  {journal}
  {\bibinfo  {journal} {Particles}\ }\textbf {\bibinfo {volume} {2}},\ \bibinfo
  {pages} {197} (\bibinfo {year} {2019})},\ \Eprint
  {http://arxiv.org/abs/1902.01089} {arXiv:1902.01089 [cond-mat.stat-mech]}
  \BibitemShut {NoStop}%
\bibitem [{\citenamefont {Akamatsu}\ \emph {et~al.}(2017)\citenamefont
  {Akamatsu}, \citenamefont {Mazeliauskas},\ and\ \citenamefont
  {Teaney}}]{Akamatsu:2016llw}%
  \BibitemOpen
  \bibfield  {author} {\bibinfo {author} {\bibfnamefont {Y.}~\bibnamefont
  {Akamatsu}}, \bibinfo {author} {\bibfnamefont {A.}~\bibnamefont
  {Mazeliauskas}}, \ and\ \bibinfo {author} {\bibfnamefont {D.}~\bibnamefont
  {Teaney}},\ }\href {\doibase 10.1103/PhysRevC.95.014909} {\bibfield
  {journal} {\bibinfo  {journal} {Phys. Rev.}\ }\textbf {\bibinfo {volume}
  {C95}},\ \bibinfo {pages} {014909} (\bibinfo {year} {2017})},\ \Eprint
  {http://arxiv.org/abs/1606.07742} {arXiv:1606.07742 [nucl-th]} \BibitemShut
  {NoStop}%
\bibitem [{\citenamefont {Schlichting}\ and\ \citenamefont
  {Teaney}(2019)}]{Schlichting:2019abc}%
  \BibitemOpen
  \bibfield  {author} {\bibinfo {author} {\bibfnamefont {S.}~\bibnamefont
  {Schlichting}}\ and\ \bibinfo {author} {\bibfnamefont {D.}~\bibnamefont
  {Teaney}},\ }\href {\doibase 10.1146/annurev-nucl-101918-023825} {\bibfield
  {journal} {\bibinfo  {journal} {Ann. Rev. Nucl. Part. Sci.}\ }\textbf
  {\bibinfo {volume} {69}},\ \bibinfo {pages} {447} (\bibinfo {year} {2019})},\
  \Eprint {http://arxiv.org/abs/1908.02113} {arXiv:1908.02113 [nucl-th]}
  \BibitemShut {NoStop}%
\bibitem [{\citenamefont {Bluhm}\ \emph {et~al.}(2020)\citenamefont {Bluhm}
  \emph {et~al.}}]{Bluhm:2020mpc}%
  \BibitemOpen
  \bibfield  {author} {\bibinfo {author} {\bibfnamefont {M.}~\bibnamefont
  {Bluhm}} \emph {et~al.},\ }\href@noop {} {\  (\bibinfo {year} {2020})},\
  \Eprint {http://arxiv.org/abs/2001.08831} {arXiv:2001.08831 [nucl-th]}
  \BibitemShut {NoStop}%
\end{thebibliography}%

\end{document}